\documentclass[%
reprint, 
superscriptaddress,
amsmath,
amssymb,
aps, 
prx,
]{revtex4-2}

\usepackage{color}

\usepackage{graphicx}
\usepackage{dcolumn}
\usepackage{bm}
\usepackage[colorlinks=true]{hyperref}
\usepackage{braket}
\usepackage{diagbox}
\usepackage{mathrsfs}
\usepackage{mathtools}
\usepackage{tabularx}

\begin{document}


\title{\textbf{Classification of topological phases of matter in stochastic systems} 
}%

\author{Dexin Li}
 \affiliation{\textit{Center for Theoretical Biological Physics, }\textit{Rice University}\textit{, Houston, Texas 77005, USA}}
 
\author{Evelyn Tang}%
 \email{Contact author: e.tang@rice.edu}
 \affiliation{\textit{Center for Theoretical Biological Physics, }\textit{Rice University}\textit{, Houston, Texas 77005, USA}}
 \affiliation{\textit{Department of Physics and Astronomy, }\textit{Rice University}\textit{, Houston, Texas 77005, USA}}

\date{\today}

\begin{abstract}
Topological phases of matter support edge states protected from random perturbations, and the identification of the discrete non-spatial symmetries and dimensions that support such phases was systematically developed for quantum systems and analogous platforms. While topological phases were also discovered in stochastic systems and posited as mechanisms for key biochemical processes, a classification consistent with their Markovian constraints remains lacking, impeding generalization to new scenarios. To address this, we develop new methods including a new homotopy approach, real-space invariants and a redefinition of the point gap to classify both transient and steady states. We find robust transient states in two symmetry classes, no symmetry and pseudo-Hermiticity respectively, while topological steady states are expected only in systems without symmetry. We identify the dimensions with non-trivial phases and their group structure, creating a rigorous framework for predicting robust behavior in active and living matter.
\end{abstract}

\maketitle
\section{Introduction}
Topological invariants identify new phases of matter that host boundary modes robustly protected against continuous deformations
~\cite{82edge, 02edge, 93chern, 06bulk}. These properties were first studied in electronic systems such as the quantum Hall effect and later extended to other quantum materials and phenomena~\cite{80QH, 82TKNN, 83homotopy, 85Topological, 85quantized, 88Haldane, 10TI, 11TI, 16band}. Since then, topological phases have been realized in several other platforms, from mechanical lattices~\cite{15phononics,18phononics, 18mechanics, 24mechanical, 26phononics}, to acoustic metamaterials~\cite{15acoustic,21acoustic, 22acoustics, 23second, 26phononics}, photonic lattices~\cite{08photonics,08reflection,09observation,14surface,18plasmonic,19photonics,23second}, electronic circuits~\cite{18circuit,19circuit}, and active matter~\cite{17curve, 20active, 22active, 26active}. To develop a systematic understanding of when topological phases could be expected, gapped Hamiltonians with discrete non-spatial symmetries were classified by symmetry and dimension~\cite{09Kitaev,16classification,19catalogue, 19complete, 19comprehensive}. As the operators governing the new platforms could be fully mapped to quantum Hamiltonians, previous classification results were directly transferable~\cite{19photonics,26phononics}. Further, since non-equilibrium phenomena and dissipation are ubiquitous in realistic settings~\cite{18edge, 18nhphases,19nhse, 20nhphysics}, classifications were also developed for non-Hermitian operators~\cite{19symm,23symm}.


Recently, topological phases have been observed in stochastic systems~\cite{25review,24ergodic,24role}, exhibiting both localized steady states~\cite{17stochastic, 18dissipative} and non-equilibrium steady states~\cite{21topology} at the boundaries. These protected modes have been proposed as biochemical mechanisms for kinetic proofreading~\cite{17stochastic}, the circadian rhythm~\cite{24mechanism}, gene transcription~\cite{25gene}, and microtubule dynamic instability~\cite{26promote}. Despite these examples, a general classification of which networks can host topological phases, similar to what exists for quantum Hamiltonians, is absent. As stochastic systems have important formal differences from quantum systems, it remains unclear if topological phases are expected in other stochastic systems, or how useful they are for biological and chemical systems generally.

Notably, stochastic systems are governed by the master equation~\cite{76network}, which has a linear operator similar to that of the Schr\"odinger equation. However, the master equation imposes new constraints of real and non-negative matrix entries, and probability conservation. These differences between quantum and stochastic systems change the respective spectra~\cite{24role,26unique}. For instance, it was proved that the bulk-boundary correspondence fails in stochastic Hermitian systems and is restored only with non-Hermitian (or non-symmetric) operators~\cite{24nonre}, necessitating a fully non-Hermitian approach for any topologically non-trivial phase unlike in quantum systems. In another difference from quantum systems, the stochastic steady state (zero mode) lies within the bulk spectrum~\cite{24role}, and thus should fall under the definition of a gapless system (e.g., similar to a quantum semimetal). However, previous works~\cite{17stochastic,18dissipative,24role} classified stochastic models using invariants for gapped systems, rendering it unclear whether methods for phases with gapped or gapless spectra are more appropriate in stochastic systems. 

Further examples illustrate the need for fundamentally new methods in the analysis and classification of stochastic topological phases. The Markovian constraints reduce the number of possible symmetry classes to 15 at maximum~\cite{25symm}, a much smaller set than the 38 possible classes in non-Hermitian quantum systems~\cite{19symm}. In addition, typical methods for the classification of quantum non-Hermitian topological phases, such as \(K\)-theoretic methods including Hamiltonian-doubling~\cite{19symm} and spectral flattening~\cite{09Kitaev}, fail in stochastic systems as they do not preserve the Markovian constraints. Most of all, there is a new matrix space which necessitates the development of a different homotopy approach.

In this paper, we analyze these challenges to obtaining a topological classification for stochastic systems and introduce new methods that circumvent them, such as a new homotopy approach for the matrix space relevant to the master equation. We also develop real-space homotopy invariants constructed from the spectral localizer~\cite{24photonic} and introduce a new definition of the point gap which enables a classification of both transient states and the steady state. For transient states, we find that only two symmetry classes remain robust: AI and BDI\(^\dagger\), which are no symmetry and pseudo-Hermiticity respectively. For the steady state, only class AI with no symmetry contains non-trivial topological phases. This completes the classification for discrete non-spatial symmetries, where we enumerate all possible phases in different dimensions. This work provides predictions for new observables and steady states in stochastic systems and consequently, in living and active matter.

\section{Challenges of a new matrix space}
\subsection{Markovian constraints in the master equation}
\label{sec:markovian}
A stochastic system is governed by the master equation \(\frac{d}{dt}\bm{p}(t)=W\bm{p}(t)\). Here, \(W\) denotes the transition matrix and \(\bm{p}(t)\) is the vector with entries \(p_i(t)\) giving the probability of the system in state \(i\) at time \(t\)~\cite{76network}. Likewise, a quantum system is governed by the Schr\"odinger equation \(\mathrm{i}\hbar\frac{d}{dt}\bm{\psi}(t)=H\bm{\psi}(t)\), where \(\bm{\psi}(t)\) is the wavefunction and \(H\) is the Hamiltonian. Since the master equation and Schr\"odinger equation are both linear dynamical equations, it is possible to partially map their respective operators. The transition matrix \(W\) in stochastic systems can be identified with the Hamiltonian \(H\) in quantum systems~\cite{24nonre,24role,21topology}, while the eigenmodes with zero eigenvalue \(\lambda_0=0\) (the steady state \(\bar{\bm{p}}\) in stochastic systems and the zero-mode in quantum systems) are analogous. This partial map has been used to calculate topological invariants in certain stochastic models~\cite{24role,21topology}.

However, the mapping remains partial because the transition matrix \(W\) is an \(n\times n\) real matrix with additional Markovian constraints 
\begin{align}
W_{ij}\geq0, i\neq j\label{Metzler},\\
\sum^n_{i=1}W_{ij}=0,\forall j\label{probcon}.
\end{align}
Constraint Eq.\eqref{Metzler} arises from the non-negativity of the transition rates from state \(j\) to state \(i\) whereas constraint Eq.~\eqref{probcon} ensures probability conservation. Since an essentially non-negative matrix is defined as a matrix satisfying the constraint Eq.~\eqref{Metzler}~\cite{94non,12matrix}, we call Eq.~\eqref{Metzler} the essentially non-negativity constraint in this work. As transition matrix \(W\) is real, it can be symmetric or non-symmetric. This distinction corresponds to Hermitian and non-Hermitian cases when complex values are allowed, such as in a quantum Hamiltonian \(H\). To facilitate comparison, we will use the most general terms of Hermitian and non-Hermitian for all matrices. 

The Markovian constraints above alter the matrix properties of \(W\) as compared to the Hamiltonian \(H\).
Nelson et al.~\cite{24nonre} proved that since the steady state of Hermitian stochastic systems is uniformly distributed, the bulk-boundary correspondence fails in the Hermitian case. This renders all Hermitian stochastic systems topologically trivial, thus any meaningful classification of stochastic systems will have to address non-Hermitian transition matrices \(W\). This exemplifies how the Markovian constraints fundamentally modify the topological classification.

The classification of stochastic systems is further plagued by issues such as the coincidence of the steady state (zero mode) with the spectrum under periodic boundary conditions (PBC)~\cite{24role}. This poses a challenge because a gap is crucial to determine if a deformation of the system is continuous~\cite{10TI,13intro}. In quantum systems, topological insulators have a conventional band gap, while the gapless spectra of phases such as semimetals~\cite{16classification,18semimetal} can exhibit a gapless point surrounded by a gapped closed surface where topological invariants can be computed. Nevertheless, while the spectra in stochastic systems are gapless and thus formally related to semimetals, recent studies~\cite{17stochastic,18dissipative,24role} suggest that they can be classified by invariants used for topological insulators. This contradiction highlights the ambiguity between phases with gapped and gapless spectra in stochastic systems. 

Another issue concerns the relation between Hermitian and non-Hermitian topological phases in stochastic systems, which does not follow existing relations in quantum systems. A non-Hermitian Hamiltonian \(H\) with a point gap can be embedded into a Hermitian Hamiltonian via \(\widetilde{H}\coloneqq\begin{pmatrix}0&H\\H^\dagger&0\end{pmatrix}\)~\cite{19symm}. \(\widetilde{H}\) preserves all the symmetries of \(H\) and an additional chiral symmetry \(\widetilde{\Sigma}=\sigma_z\otimes\mathbb{I}_n\) (where \(\sigma_z\) is the third Pauli matrix and \(\mathbb{I}_n\) is the \(n\times n\) identity matrix). Consequently, the topological classification of non-Hermitian gapped Hamiltonians is essentially that of Hermitian ones with an additional chiral symmetry. Now, one could also define \(\widetilde{W}\coloneqq\begin{pmatrix}0&W\\W^T&0\end{pmatrix}\) in stochastic systems. However, this would break the constraints in Eqs.~\eqref{Metzler} and \eqref{probcon} as the off-diagonal elements of \(\widetilde{W}\) would not be non-negative nor would the column sums of \(W^T\) vanish. Hence, \(\widetilde{W}\) will no longer represent a transition matrix. It is therefore necessary to find new methods for the classification of both Hermitian and non-Hermitian stochastic systems.

\subsection{Failure of \(K\)-theory and hence classification directly in \(W\) space}
\label{failureK}
The comprehensive topological classification of gapped Hermitian quantum systems with discrete non-spatial symmetries was first developed by Kitaev via a \(K\)-theoretic framework~\cite{09Kitaev} and soon after, Ryu et al. showed how to construct the corresponding Dirac Hamiltonians and topological invariants via dimensional reduction~\cite{10dimensional}. Specifically, Kitaev proved that in the \(q\)-th symmetry class, it is possible to construct the classifying space \(\mathcal{R}_{q}\) (\(\mathcal{C}_{q}\)). It follows that the classification in \(d\) dimensions reduces to the zeroth homotopy set \(\pi_0(\mathcal{R}_{q-d})\) (\(\pi_0(\mathcal{C}_{q-d})\)) which counts its connected components. This approach, rooted in the \(K\)-theory of vector bundles equipped with a Clifford algebra structure~\cite{14bloch,09Kitaev}, has led to a topological classification for quantum systems in the ten classes~\cite{09Kitaev}. However, extending this classification to stochastic systems is challenging because of the contradiction between the essentially non-negativity constraint Eq.~\eqref{Metzler} and invertibility, which will be explained below. 

Firstly, the spectral flattening technique is used in quantum systems with spectral decomposition to explicitly construct continuous deformation paths, but this method is not applicable to stochastic systems. More specifically, one can obtain a flattened Hamiltonian \(Q\) satisfying \(Q^2=\mathbb{I}_n\) continuously from the original Hamiltonian \(H\)~\cite{09Kitaev,19symm}. In the stochastic case, in contrast, it is impossible to construct the corresponding flattened \(Q\) from a transition matrix \(W\) because \(W\) is not invertible due to the zero eigenvalue. Although this issue can be addressed by tilting \(W\) away from 0 to render it invertible~\cite{17stochastic, 24role}, such tilting does not work in 0D. Moreover, even if \(W\) is tilted to be invertible in higher dimensions, the Markovian constraints hold only under a particular representation with this method, making it unclear how to apply these constraints in other bases more generally (details in Appendix~\ref{sec:failflat}). The inapplicability of spectral flattening in stochastic systems precludes the general use of standard K-theoretic methods, such as the association with the Clifford algebra structure~\cite{09Kitaev}. 

Secondly, in translationally invariant quantum systems, topological classification is typically established in the momentum-space ($\bm{k}$-space) picture~\cite{09Kitaev,14bloch,16band,19symm}. Similarly, in stochastic cases, it would be natural to consider the reciprocal-space transition matrix \(W(\bm{k})\) obtained from a transition matrix \(W\) via Fourier transformation. 
However, there is a significant difference between quantum and stochastic cases. In quantum systems, given a general matrix \(H(\bm{k})\), we can use the method of equivariant homotopy~\cite{14bloch} to directly determine the topological properties of \(H(\bm{k})\). Here \(H(\bm{k})\) can be regarded as a specific periodic matrix-valued function of \(\bm{k}\) rather than the Fourier form of a Hamiltonian \(H\). However, this method does not work for Markovian constraints which cannot be represented by a simple linear/anti-linear operator (details in Appendix~\ref{sec:mismatch}). Hence, given a general matrix \(W(\bm{k})\) in stochastic systems, we have to first determine whether its inverse Fourier form \(W\) is indeed a transition matrix satisfying Markovian constraints Eqs.~\eqref{Metzler} and \eqref{probcon}. Consequently, in stochastic systems, it is natural to first study \(W\) and then its Fourier form \(W(\bm{k})\).

To address these issues and identify the topological features of stochastic systems, we will study the topological classification of stochastic systems directly in \(W\) space rather than in reciprocal space. More concretely, we use the original representation of \(W\) (which may describe an abstract or configuration space~\cite{21topology,24role}), similar to a real-space approach in quantum systems. Specifically, we will establish the topological classification by homotopy and construct invariants for \(W\) using the spectral localizer proposed for real-space topology~\cite{24photonic,24realspace}. With such real-space invariants, a Fourier transformation of \(W\) can be made to use those invariants in \(\bm{k}\)-space. We provide an explicit example of this procedure in Appendix~\ref{sec:onedwind}.

\begin{figure*}[htbp]
\centering
\includegraphics[scale=0.31]{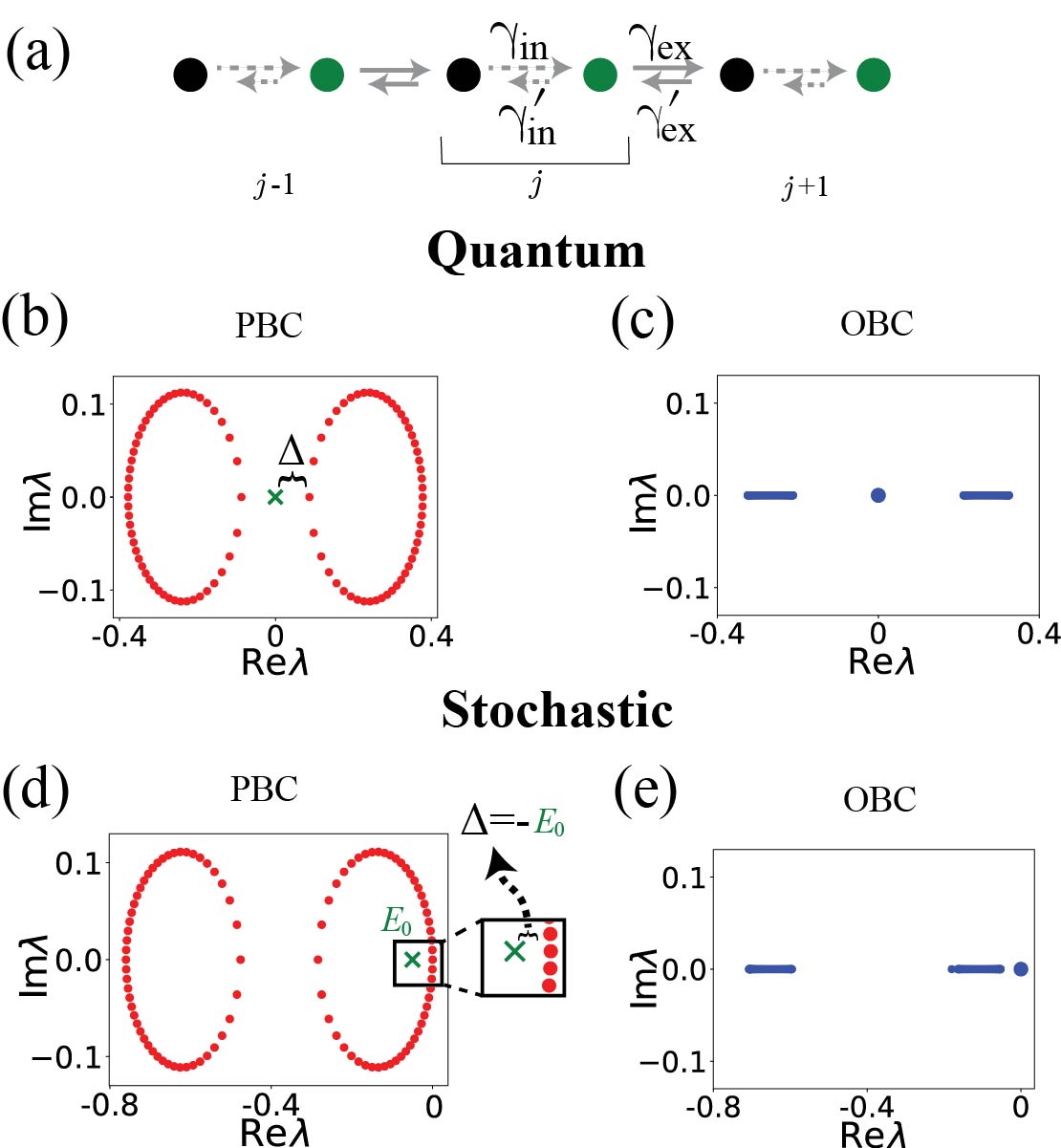}
\caption{(a) Schematic of the non-Hermitian SSH model, defined in Eq.~\eqref{Eq:SSH}. (b,c) The spectra \(\lambda\) of quantum systems are shown in red for periodic boundary conditions (PBC) and in blue for open boundary conditions (OBC). The zero mode exists under OBC and disappears under PBC, which allows a straightforward definition of a bulk gap \(\Delta\) and the point gap denoted by a green cross. (d,e) Similarly, the spectra \(\lambda\) of stochastic systems are shown in red for periodic boundary conditions (PBC) and in blue for open boundary conditions (OBC). In contrast to quantum systems, the steady state with zero eigenvalue appears under both PBC and OBC, which illustrates the difficulty of defining a bulk gap due to the lack of one in PBC. To address this, we introduce a modified point gap by shifting it slightly away from zero. Inset of (d): the green cross is set at \(E_0\) rather than at zero as in (b) to ensure that it does not coincide with the eigenvalue zero, leading to a non-zero bulk gap \(\Delta=-E_0\). The parameters used are \(\gamma_{\mathrm{in}}=0.16\), \(\gamma_{\mathrm{ex}}=0.4\), \(\gamma_{\mathrm{in}}'=0.02\), \(\gamma_{\mathrm{ex}}'=0.18\) and \(n=70\).}
\label{fig:gap}
\end{figure*}

\section{Only two classes remain robust}
Before establishing the topological classification, we first identify all possible symmetry classes that satisfy the Markovian constraints. Lucas S\'a et al.~\cite{25symm} showed that since transition matrices are real, the number of classes in non-Hermitian stochastic systems is reduced to 15 from 38 in quantum systems~\cite{19symm} as now only three symmetry generators, represented as real invertible matrices \(C_+\), \(C_-\) and \(S\), remain. They defined these generators by the commutation or anticommutation relations with \(W'\coloneqq W-\frac{\mathrm{Tr}W}{n}\mathbb{I}_n\), where the transition matrix \(W\) is offset by a multiple of the identity to obtain a traceless matrix \(W'\). Specifically, \(C_+\) and \(C_-\) satisfy
\begin{align}
C_+ W'^TC_+^{-1} =W',\\
C_- W'^TC_-^{-1} =-W',
\label{cminus}
\end{align}
while \(S\) satisfies
\begin{align}
SW'S^{-1}=-W'.
\label{subl}
\end{align}

Closer inspection reveals that not all of these symmetry classes will host topologically non-trivial phases. This is because topological properties should be insensitive to random perturbations to \(W\) that preserve its symmetries~\cite{13intro}. However, the original \(W\) does not anticommute with \(C_-\) and \(S\); only \(W'\) does. As a result, perturbations to W would require another offset of the identity to remain traceless and for the anticommutation relations to be preserved. Notably, generic perturbations without this fine-tuning will break these symmetries, and hence they are not robust to perturbations. This explains why a previous model that satisfied \(S\) symmetry in Eq.~\eqref{subl} was successfully analyzed with a topological invariant for the symmetry class without \(S\)~\cite{24role}. Hence, we will not consider symmetry generators that only work on \(W'\) but fail on the actual \(W\) operator. 

For robust topological phases, we are left with only the symmetry generator \(C_+\) which satisfies
\begin{align}
C_+W^TC_+^{-1}=W,
\label{trs}
\end{align}
and allows the three symmetry classes: AI (no \(C_+\)), BDI\(^\dagger\) (\(C_+=C_+^{T}\)) and DIII\(^\dagger\) (\(C_+=-C_+^{T}\)). We focus on the case of ergodic systems, i.e., where a state can be reached from any other state in a finite number of steps, as these are the typical cases~\cite{97markov}. Notably, ergodic systems have a unique steady state due to the Perron-Frobenius theorem~\cite{12matrix}. Since DIII\(^\dagger\) is subject to a non-Hermitian generalization of Kramers theorem, the spectrum is two-fold degenerate~\cite{19symm,25symm}. As this does not fall into the framework of ergodic systems, we will not discuss this symmetry in this paper, although the methods we develop can be extended to non-ergodic systems in future work.

To summarize, non-Hermitian ergodic stochastic systems have only two topologically robust classes, namely, AI and BDI\(^\dagger\). Class AI has no symmetries (besides the condition of real transition rates) while class BDI\(^\dagger\) contains a generalized bipartite structure which will be discussed in Sec.~\ref{sec:BDIdagger}. 
In the Hermitian limit, classes BDI\(^\dagger\) and DIII\(^\dagger\) cannot exist, so there remains only class AI. 

\section{Generalization of point gap in stochastic systems}
\subsection{Point gap as a tunable probe}
\label{pointgap}
Before we can analyze the topological phases present in the two remaining symmetry classes for stochastic systems, AI and BDI\(^\dagger\), we first have to address issues regarding gaplessness of the spectrum since the eigenvalue of the steady state is fixed at 0 as introduced in Sec.~\ref{sec:markovian}. To start, we introduce some useful concepts about the gap. To quantitatively distinguish between gapped and gapless systems, it is convenient to define a bulk gap \(\Delta\) as the minimum distance between a chosen reference point \(E_0\) and the spectrum of a given system. With this definition, a system is gapped at \(E_0\) if \(\Delta>0\) and gapless if \(\Delta=0\). In the \(\Delta>0\) case, the reference point \(E_0\) can be defined as a point gap~\cite{19symm}, which can probe the topology around that point in the spectrum~\cite{24photonic}. By choosing \(E_0\) properly, it is possible to establish the topological classification around it. 

However, this definition is problematic in stochastic systems, which can be shown using the example of the one-dimensional (1D) non-Hermitian Su–Schrieffer–Heeger (SSH) model~\cite{24nonre}; see Fig.~\ref{fig:gap}(a). This is defined as
\begin{align}
\begin{split}
\hat{H}=\sum^n_{j=1}(\gamma_{\mathrm{in}}\ket{j, B}\bra{j, A}+\gamma'_{\mathrm{in}}\ket{j, A}\bra{j, B})\\
	+\sum^{n-1}_{j=1}(\gamma_{\mathrm{ex}}\ket{j+1, A}\bra{j, B}+\gamma'_{\mathrm{ex}}\ket{j, B}\bra{j+1, A})
\end{split}
\label{Eq:SSH}
\end{align}
on a 1D network with \(n\) cells where there are two states \(\ket{j, A}\) and \(\ket{j, B}\) in the \(j\)-th unit cell, indicated by black and green circles respectively in Fig.~\ref{fig:gap}(a). Here \(\gamma_{\mathrm{in}}\) and \(\gamma_\mathrm{ex}\) are the forward rates: internal rates within the unit cell and external rates between the unit cells respectively, while \(\gamma_\mathrm{in}'\) and \(\gamma_\mathrm{ex}'\) are the corresponding backward rates. This model can describe a tight-binding Hamiltonian in a quantum system where arrows represent hoppings between sites, or the off-diagonal part of a transition matrix in a stochastic system where arrows represent transitions between states.

For quantum systems, we construct the corresponding matrix \(H\) under periodic boundary conditions (PBC) and open boundary conditions (OBC). As illustrated in Fig.~\ref{fig:gap}(b), the PBC spectrum marked in red is gapped at 0, indicated by the green cross, while topologically protected zero modes can appear in the OBC spectrum marked in blue, as illustrated in Fig.~\ref{fig:gap}(c). Thus it is natural to choose \(E_0=0\) with \(\Delta>0\). For stochastic systems, we construct the transition matrix \(W=H-D\) under PBC and OBC respectively with \(D_{ij}=\delta_{ij}\sum_{k\neq j}H_{kj}\). Since we are interested in the steady state with \(\lambda_0=0\), \(E_0\) should be chosen relative to zero, which is always an eigenvalue of any stochastic system (due to probability conservation Eq.~\eqref{probcon}~\cite{76network}) and can be observed in Fig.~\ref{fig:gap}(d,e).

Hence, we can set the point gap \(E_0\) of \(W\) as a small negative number denoted by a green cross in Fig.~\ref{fig:gap}(d). To connect \(E_0\) to the steady state, we regard the point gap as a tunable probe that detects whether the rest of the spectrum can continuously extend to 0 along the real axis. More concretely, we can check the distance between 0 and the greatest non-zero real eigenvalue. If this quantity is non-zero, we can always find a negative number \(E_0\) such that no eigenvalues intersect the interval \([E_0,0)\). It is thus possible to re-define the bulk gap \(\Delta\) as the distance between \(E_0\) and \(0\), i.e., \(\Delta=-E_0\), as indicated in the inset of Fig.~\ref{fig:gap}(d). Accordingly, a stochastic system with \(\Delta>0\) is defined as a gapped system. In this case, all arguments for gapped systems in quantum cases can be applied~\cite{10TI,16classification}. Otherwise, if the spectrum extends continuously from 0 along the negative semi-axis, it is impossible to find such an \(E_0\), and hence the corresponding system is defined to be gapless. This gapless case will be discussed later in Secs.~\ref{dDHermitian} and~\ref{onedAI}.

Besides the point gap, there is another type of gap in non-Hermitian systems: the line gap~\cite{19symm}. However, the line gap might be of limited use in stochastic systems as only the real line gap can be defined in stochastic systems. It was shown that a real line gap can render quantum systems effectively Hermitian~\cite{19symm}. Applying this to stochastic systems would result in only topologically trivial phases, as previously shown in \cite{24nonre} (and discussed in Section \ref{sec:markovian}). Therefore, the classification of systems with a line gap will be stated as natural generalizations of the Hermitian result in Secs.~\ref{sec:0dgeneral} and \ref{dDHermitian}.

\subsection{Equivalent framework: Shifted transition matrix}
While the point gap \(E_0\) as defined in the last section addresses the issues regarding gaplessness of the spectrum, using this framework would require the calculation of the point gap using \(\operatorname{det}{(W-E_0\mathbb{I}_n)}\). In fact, \(E_0\) is a point gap of the transition matrix \(W\) if and only if the spectrum of \(W\) does not contain \(E_0\), i.e., when \(\operatorname{det}{(W-E_0\mathbb{I}_n)}\neq0\). This is a condition that must be checked under continuous deformations of $W$ and is rather tedious. 

To relax this requirement, we move to an equivalent picture leaving the point gap at 0 but shifting the whole spectrum by \(-E_0\). As this just re-defines the origin compared to the previous picture, the two frameworks are equivalent in avoiding the gaplessness of the spectrum but this second approach is more convenient. 

To see this, we define a matrix \(W^{\mathrm{sh}}\) by applying a constant spectral shift to \(W\), i.e.,
\begin{align}
W^{\mathrm{sh}}\coloneqq W-E_0\mathbb{I}_n.
\label{shiftedW}
\end{align}
Now, 0 is not an eigenvalue of \(W^{\mathrm{sh}}\) if and only if \(E_0\) is not an eigenvalue of \(W\), exactly the case where the topological classification is well-defined. The point-gapped condition \(\operatorname{det}{(W-E_0\mathbb{I}_n)}\neq0\) is now just that \(W^{\mathrm{sh}}\) should be invertible. This more relaxed condition is easier to check during any continuous deformation or matrix manipulation. 

Therefore, we will move to studying the shifted transition matrix \(W^{\mathrm{sh}}\) in what follows. \(W^{\mathrm{sh}}\) still satisfies constraint Eq.~\eqref{Metzler} whereas constraint Eq.~\eqref{probcon} is now modified to
\begin{align}
\sum^n_{i=1}W^{\mathrm{sh}}_{ij}=-E_0,\forall j,
\label{shiftedcon}
\end{align}
i.e., for each \(j\), the sum over the \(j\)-th column has a constant shift as well.

Then, the classification of the steady state of \(W\) is recovered by taking the limit 
\begin{align}
\lim_{E_0\to0^-}W^{\mathrm{sh}}\equiv\lim_{E_0\to0^-}(W-E_0\mathbb{I}_n)=W.
\end{align}
For the classification of transient states (states each associated with a non-zero eigenvalue), we can also keep \(E_0\) fixed at a non-zero value. 

Note that another way around the gaplessness of the spectrum is to introduce a tilted transition matrix \(W^\lambda\)~\cite{17stochastic, 24role} as mentioned in Sec.~\ref{failureK}. As shown in Appendix~\ref{sec:tilt}, this tilting approach is equivalent to our approach in one dimension. However, the tilting approach cannot be applied in 0D since it has to vary with position in state space. In contrast, our approach provides a systematic framework for all dimensions and is computationally efficient.

\section{Homotopy approach}
\subsection{Homotopy is crucial for topological classification}
\label{homotopy}
Now we turn to a homotopy approach to establish the topological classification. Before we dive into the details, we first introduce the main idea of this approach. Two \(n\times n\) matrices are in the same topological phase if and only if they can be continuously deformed into one another while preserving certain symmetries~\cite{10TI}. Here, continuous deformations are mathematically characterized by the homotopy type of the space of certain matrices and as a result, the topological classification is determined by homotopy classes of these matrix spaces. In quantum systems, the spaces of certain Hamiltonians are homotopy equivalent to the classifying spaces in \(K\)-theory in the limit of \(n\to\infty\)~\cite{09Kitaev}. According to these homotopy equivalences, standard classification results in \(K\)-theory can be used to establish the classification of quantum systems. 

In stochastic systems, the topological classification also depends on the homotopy types of the relevant matrix spaces, i.e., the spaces of shifted transition matrices defined in Eq.~\eqref{shiftedW} for each symmetry class. However, as explained previously (Sec.~\ref{failureK}), standard \(K\)-theory cannot be applied to stochastic systems. Hence we need to develop a new homotopy approach for the space of shifted transition matrices, where the Markovian constraints Eqs.~\eqref{Metzler} and \eqref{shiftedcon} alter their properties as compared to quantum Hamiltonians (which can be constructed from classical Lie groups). 

For instance, consider non-Hermitian systems in class AI (with no symmetry). The matrix space in the quantum case is the well-known Lie group GL\((n,\mathbb{R})\) while the matrix space in the stochastic case is its subspace \(\mathcal{W}(n,-E_0)\) of all \(n\times n\) invertible shifted transition matrices \(W^{\mathrm{sh}}\) as defined in Eq.~\eqref{shiftedW}, which is not even a group. This can be seen by considering that since transition matrices have negative entries (in the diagonal), the product of two transition matrices may not satisfy the essentially non-negativity constraint Eq.~\eqref{Metzler} and hence \(\mathcal{W}(n,-E_0)\) is not closed under multiplication. Overall, little is known about the homotopy of the space of transition matrices. 

In what follows, we develop a proof of the homotopy equivalence between the space of shifted transition matrices and a classifying space, in order to obtain a topological classification.

\subsection{Homotopy of transition matrix space}
\label{sec:non-negative}
In this section, we show homotopy equivalences between different matrix spaces that permit a topological classification. We start with two matrix spaces for \(W^{\mathrm{sh}}\): a space \(\mathcal{W}(n,-E_0)\) defined in the last section for non-Hermitian cases, and its symmetric subspace \(\mathcal{W}^{\mathrm{sym}}(n,-E_0)\) for Hermitian cases. Specifically, the symmetric subspace of a space \(\mathcal{X}\) can be defined as \(\mathcal{X}^\mathrm{sym}\coloneqq\{X\in\mathcal{X}\mid X=X^T\}\).

We will demonstrate a correspondence between the topological classification for \(W^{\mathrm{sh}}\) and the classification of quantum cases, by proving the homotopy equivalence between \(\mathcal{W}(n,-E_0)\simeq \mathrm{GL}(n-1,\mathbb{R})\) and \(\mathcal{W}^{\mathrm{sym}}(n,-E_0)\simeq \mathrm{GL}^{\mathrm{sym}}(n-1,\mathbb{R})\). This equivalence can be intuitively understood by examining the Markovian constraints Eqs.~\eqref{Metzler} and \eqref{probcon}, as follows. The essentially non-negativity constraint Eq.~\eqref{Metzler} can be removed from a homotopy perspective (details in Appendix~\ref{sec:unified}). On the other hand, the probability conservation Eq.~\eqref{probcon} reduces the effective order of the transition matrix by 1 since the fixed column sums of an \(n\times n\) transition matrix impose that each column vector depends on \(n-1\) degrees of freedom. This results in the topological classification of stochastic systems of size \(n\) being the same as that of quantum systems of size \(n-1\). Furthermore, this argument holds in line-gapped stochastic systems as well, consolidating our classification framework for both point-gapped and line-gapped scenarios.

Hence, in both Hermitian and non-Hermitian cases, these homotopy equivalences imply that the topological properties of an \(n\times n\) shifted transition matrix in stochastic systems are similar to those of an \((n-1)\times (n-1)\) Hamiltonian in quantum systems. This conclusion is consistent with previous results showing that the stochastic spectrum for a system of size \(n\) overlaps with that of a similar quantum system of size \(n-1\)~\cite{26unique}. 

Next, as proved in Appendices~\ref{sec:homotopynonh} and \ref{sec:homotopyh}, the spaces that we consider in this work are further homotopy equivalent to the space of non-negative matrices~\cite{26homotopy,26sym} and stochastic matrices, which describe the discrete-time Markovian classical stochastic process~\cite{26discrete}. Hence even though we discuss the topological classification and construct the corresponding topological invariants for transition matrices in the continuous-time equivalent, this classification framework can be directly applied to such discrete-time processes. This explains why in the discrete-time process characterized by a stochastic matrix, the topological invariants defined in the quantum case are still valid.

\subsection{Finite point gap classifies transient states and discrete-time stochastic processes}
\label{subsec:homotopy}
Our establishment of the above homotopy equivalences now permits the calculation of the homotopy groups of the matrix spaces --- or more simply, the classifying spaces. Without loss of generality, we will start with the classification of non-Hermitian stochastic systems since the Hermitian case proceeds similarly. We first construct the space \(\mathcal{W}_q(n,-E_0)\) consisting of all \(W^{\mathrm{sh}}\) in the \(q\)-th symmetry class. Since the topological classification is established in the thermodynamic limit \(n\to\infty\)~\cite{16classification}, the classifying space is \(\mathcal{W}_q(-E_0)\coloneqq\mathcal{W}_q(\infty,-E_0)\). 

The zero-dimensional (0D) topological classification for transient states of non-Hermitian stochastic systems in the \(q\)-th class is given by \(\pi_0(\mathcal{W}_{q}(-E_0))\), analogous to the quantum one obtained from the corresponding classifying space~\cite{09Kitaev}. For example, take class AI. Since there is no additional symmetry constraint in this class, the matrix space is just the most general one defined in Sec.~\ref{homotopy}, i.e., \(\mathcal{W}_\textrm{AI}(n,-E_0)=\mathcal{W}(n,-E_0)\) and the corresponding classifying space is \(\mathcal{W}_{\mathrm{AI}}(-E_0)\equiv\mathcal{W}(\infty,-E_0)\simeq\mathrm{GL}(\infty,\mathbb{R})\simeq\mathcal{R}_1\), which is the classifying space of non-Hermitian quantum systems in class AI. The 0D topological classification is then given by \(\pi_{0}(\mathcal{W}_{\mathrm{AI}}(-E_0))=\pi_0(\mathcal{R}_1)\cong\mathbb{Z}_2\). This means that in 0D cases, there are only two phases, i.e., either topological or trivial. Thus, for transient states, the topological classification is exactly the same as in the quantum case and we can obtain the topological invariant for \(W^{\mathrm{sh}}\) analogously. 

This approach can be extended to higher dimensions: in \(d\) dimensions (\(d\)-D), we establish the topological classification using \(d\)-D real-space homotopy invariants~\cite{24photonic}. Now, the shifted transition matrix \(W^{\mathrm{sh}}\) and position matrices \(X_i\) \((i=1,\dots, d)\) are combined to construct the spectral localizer. Since these real-space invariants depend only on the homotopy class, the \(d\)-D topological classification for transient states is the same as in the quantum case.

Moreover, even though a non-zero point gap can only correspond to transient states in systems governed by a continuous-time classical stochastic process, as discussed in Sec.~\ref{sec:non-negative}, the classification and the corresponding topological invariants can be applied to other systems. This includes a generic point gap in a discrete-time classical stochastic process, which can predict the existence of nonvanishing flow of the steady state~\cite{26discrete}. This comprehensive topological classification for systems with a non-zero point gap is therefore of relevance to other settings beyond the ones we discuss.

\subsection{Taking the limit of the point gap for the steady state}
\label{subsec:homotopy_steady}
As the steady state is typically of greatest interest, we will focus on this state by taking the limit \(E_0\to0^-\) of \(W^{\mathrm{sh}}\). In 0D, we further examine \(\pi_0(\mathcal{W}_q)\equiv\pi_0(\lim_{E_0\to0^-}\mathcal{W}_q(-E_0))\). Specifically, we will use a topological invariant first defined in quantum systems, the spectral localizer~\cite{24photonic}, to count the number of connected components of the matrix space of \(W^{\mathrm{sh}}\). Then the classification of the steady state is given by that invariant for \(W^{\mathrm{sh}}\equiv W-E_0\mathbb{I}_n\) in the limit of \(E_0\to0^-\). 

While we can use such \(d\)-D topological invariants defined in quantum systems to characterize the topology of stochastic systems, these real-space invariants are formally constructed in a more abstract way. In contrast, \(\bm{k}\)-space invariants often better illustrate the physics. For instance, \(\bm{k}\)-space invariants provide an intuitive perspective for explaining the classification in the limit of \(E_0\to 0^-\), as we will show in the following sections. Hence, once the real-space topological invariants are determined, they can be used to obtain the corresponding \(\bm{k}\)-space forms for \(W^{\mathrm{sh}}(\bm{k})\), similar to these invariants in quantum systems~\cite{24realspace}. Consequently, in the steady-state limit of \(E_0\to0^-\), we will assume that translation invariance holds and use both real-space and \(\bm{k}\)-space invariants to classify the steady state for \(W^{\mathrm{sh}}(\bm{k})\equiv W(\bm{k})-E_0\mathbb{I}_n\).

Note that even though we consider \(\bm{k}\)-space topological invariants in translationally invariant systems for simplicity to obtain the classification results in the steady-state limit, the real-space invariants from the spectral localizer are constructed for more general cases without translation invariance and for both transient states and the steady state.

\section{Classification of transient states}
\label{sec:classtransient}
We found in the previous section~\ref{subsec:homotopy} that for transient states, the classifications of quantum and stochastic systems are identical in class AI. Since there is only class AI in Hermitian stochastic systems, we can directly list this classification of the transient states in all Hermitian stochastic systems in the upper part of Table~\ref{tabletransientH}. For non-Hermitian stochastic systems, since the topological classification of class AI is also the same as the quantum one as discussed in Sec.~\ref{subsec:homotopy}, we can directly list those classes in the lower part of Table~\ref{tabletransientH}, first row.

In the rest of this section, we will focus on class BDI\(^\dagger\). Here, we will establish the topological classification of stochastic systems in class BDI\(^\dagger\) via the correspondence between the homotopy of class BDI\(^\dagger\) and that of class AI. We first determine the explicit formula for \(W^{\mathrm{sh}}\) in class BDI\(^\dagger\). In BDI\(^\dagger\), the corresponding generator can be chosen as \(C_+=\sigma_x\otimes\mathbb{I}_n\) where \(\sigma_x\) is the first Pauli matrix. Hence the shifted transition matrix can be written as
\begin{align}
W^{\mathrm{sh}}=\begin{pmatrix}
A&B\\C&A^T
\end{pmatrix},
\end{align}
where both \(B\) and \(C\) are symmetric matrices. All such invertible matrices \(W^{\mathrm{sh}}\) that satisfy the constraints Eqs.~\eqref{Metzler} and \eqref{shiftedcon} form a matrix space \(\mathcal{W}_\textrm{BDI\(^\dagger\)}(2n,-E_0)\).

Here, the symmetry generator \(C_+\) represents pseudo-Hermiticity~\cite{19symm}, which can also be interpreted as a generalized detailed balance since a transition matrix satisfying detailed balance must possess \(C_+\)~\cite{25symm}. An example of a stochastic system in class BDI\(^\dagger\) can be found in a generalized bipartite graph where hopping between nodes of the same type is allowed. This is restricted by the condition that the transition rate within one part of the graph is equal to the corresponding reverse transition rate in the other part~\cite{25symm}. 

Next we define a new matrix 
\begin{align}
D=W^{\mathrm{sh}} C_+=\begin{pmatrix}
B&A\\A^T&C
\end{pmatrix}.
\end{align}
Since \(C_+\) is an invertible matrix, this linear transformation induces a homeomorphism between the space \(\mathcal{W}_\textrm{BDI\(^\dagger\)}(2n,-E_0)\) of \(W^{\mathrm{sh}}\) and the space of \(D\). Consequently, these two spaces are of the same homotopy type. As in the case of class AI discussed in Sec.~\ref{sec:homotopynonh}, we can relax the shifted Markovian constraints Eqs.~\eqref{Metzler} and \eqref{shiftedcon} to the non-negativity constraint 
\begin{align}
W^{\mathrm{sh}}_{ij}\geq0,\forall i, j.
\label{non-negativity}
\end{align}

Since an arbitrary \(2n\times 2n\) invertible non-negative symmetric matrix can always be expressed as a matrix of the form \(D\), which is symmetric due to \(B\) and \(C\) being symmetric, the corresponding classifying space \(\mathcal{W}_\textrm{BDI\(^\dagger\)}(-E_0)\) is exactly \(\mathcal{R}_0\), which is the classifying space of Hermitian quantum systems in class AI. Thus, the classifying space of transient states in class BDI\(^\dagger\) for non-Hermitian stochastic systems is the same as that of class AI Hermitian quantum systems.

Therefore, we showed that the classification of transient states of non-Hermitian stochastic systems in both class AI and BDI\(^\dagger\) is the same as that of quantum systems~\cite{19symm}. Since this exhausts all possible symmetry classes, we list the full classification results of non-Hermitian stochastic systems in the lower part of Table~\ref{tabletransientH}.

\begin{table}[htbp]
\caption{Topological classification for transient states (with non-zero eigenvalues) of all stochastic systems. Only class AI with no symmetry survives the constraints on a transition matrix in the Hermitian case (upper part). In the non-Hermitian case, two symmetry classes AI and BDI\(^\dagger\), with pseudo-Hermiticity, survive (lower part). Possible phases are enumerated by spatial dimension \(d\); since the period of dimensions is 8 as in quantum systems, only dimensions up to 7D are listed. Entries give the group of topological invariants available in each case: $0$ where only the trivial phase exists, and \(\mathbb{Z}_2\), \(2\mathbb{Z}\) or \(\mathbb{Z}\) otherwise.}
	\begin{tabular}{|c|c|c|c|c|c|c|c|c|}
		\hline
		\hline
		\diagbox[height=1.5em]{Class}{\quad\(d\)}&\;0\; &\;1\; &\;2\; &\;3\; &\;4\; &\;5\; &\;6\; &\;7\;\\
\hline
\multicolumn{9}{|c|}{\textbf{Hermitian systems}}\\
\hline
		AI&\(\mathbb{Z}\)&0 &0 &0 &\(2\mathbb{Z}\) &0 &\(\mathbb{Z}_2\) &\(\mathbb{Z}_2\)\\
\hline\hline
\multicolumn{9}{|c|}{\textbf{Non-Hermitian systems}}\\
\hline
AI & \(\mathbb{Z}_2\) & \(\mathbb{Z}\) & 0 & 0 & 0 & \(2\mathbb{Z}\) & 0 & \(\mathbb{Z}_2\)\\
\hline
BDI\(^\dagger\) & \(\mathbb{Z}\) & 0 & 0 & 0 & \(2\mathbb{Z}\) & 0 & \(\mathbb{Z}_2\) & \(\mathbb{Z}_2\)\\
\hline
\end{tabular}
\label{tabletransientH}
\end{table} 

\begin{figure}[htbp]
\centering
\includegraphics[scale=0.3]{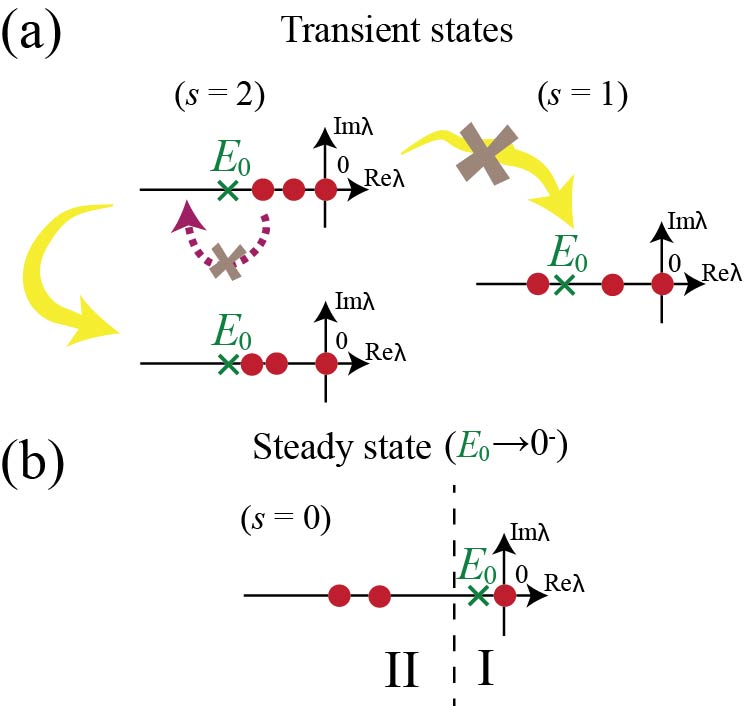}
\caption{Restrictions on the motion of eigenvalues \(\lambda\) of \(W\) (red dots) in Hermitian 0D stochastic systems for (a) transient states and (b) the steady state. The topological invariant is defined as \(s\coloneqq n_+-1\), with \(n_+\) the number of eigenvalues above the point gap \(E_0\) (green cross). (a) With \(E_0\neq0\), the phases with \(s=2\) can be continuously deformed into each other (yellow arrow). However, they cannot be continuously deformed into the phase with \(s=1\) (blocked arrow) because Hermiticity restricts eigenvalues to move only along the real axis, preventing them from passing through \(E_0\). (b) In the steady-state limit (\(E_0\to0^-\)), only the phase with \(s=0\) remains and hence must be the trivial phase. The dashed line between \(E_0\) and the first non-zero eigenvalue divides the plane into regimes I and II. Since regime I is the same in all ergodic systems and there is no point gap in regime II, all systems can be continuously deformed into each other.}
\label{fig:zerodimh}
\end{figure}

\section{Classification of the steady state}
\subsection{Hermitian stochastic systems are trivial}

\subsubsection{0D class AI}
\label{sec0daih}

While transient states inherit the classification from quantum systems in class AI as shown above, the steady state will require development of new results from obtaining the limit of the point gap. We start with Hermitian 0D stochastic systems in class AI, where we can derive the topological invariant from \(n_+(W^{\mathrm{sh}})\), the number of positive eigenvalues of \(W^{\mathrm{sh}}\in\mathcal{W}_{\mathrm{AI}}^{\mathrm{sym}}(n,-E_0)\)~\cite{24photonic}. To show this, we first demonstrate that \(n_+(W^{\mathrm{sh}})\) is constant on each connected component of \(\mathcal{W}_{\mathrm{AI}}^{\mathrm{sym}}(n,-E_0)\), i.e., the subspace where any two matrices can be connected by a continuous path. 

To be more specific, we consider \(W^{\mathrm{sh}}_0\), \(W^{\mathrm{sh}}_1\in\mathcal{W}_\textrm{AI}^{\mathrm{sym}}(n,-E_0)\) that can be continuously deformed into each other via a path in \(\mathcal{W}_\textrm{AI}^{\mathrm{sym}}(n,-E_0)\). If \(n_+(W^{\mathrm{sh}}_0)>n_+(W^{\mathrm{sh}}_1)\) holds, some positive eigenvalues of \(W^{\mathrm{sh}}_0\) have to pass through zero during a continuous deformation since any eigenvalue \(\lambda(W^{\mathrm{sh}})\) is a continuous real function of \(W^{\mathrm{sh}}\). However, such a deformation breaks the invertibility of the matrix and is therefore not a continuous path in \(\mathcal{W}_\textrm{AI}^{\mathrm{sym}}(n,-E_0)\). 

We show this intuitively using an example in the original space of \(W\) with the shifted point gap at \(E_0\) (green cross) described in Sec.~\ref{pointgap} (see Fig.~\ref{fig:zerodimh}(a)). The space of all possible \(W\)'s in Hermitian 0D stochastic systems in class AI is homeomorphic to \(\mathcal{W}_{\mathrm{AI}}^{\mathrm{sym}}(n,-E_0)\). Hence, \(n_+(W^{\mathrm{sh}})\) can also be regarded as the number of eigenvalues of \(W\) above \(E_0\). Since eigenvalues of \(W\) (red dots) can only be moved along the real axis without passing through \(E_0\) (indicated by the dashed arrow), \(W\) with \(n_+=3\) cannot be continuously deformed into \(W\) on the right with \(n_+=2\).

Therefore, if \(W^{\mathrm{sh}}_0\) and \(W^{\mathrm{sh}}_1\) are located in the same connected component of \(\mathcal{W}_{\mathrm{AI}}^{\mathrm{sym}}(n,-E_0)\), the equality \(n_+(W^{\mathrm{sh}}_0)=n_+(W^{\mathrm{sh}}_1)\) holds. In addition, each connected component has a distinct fixed value of \(n_+(W^{\mathrm{sh}})\), and hence \(n_+(W^{\mathrm{sh}})\) is a topological invariant that completely characterizes the system topology. When \(n\) is fixed, since \(n=n_+(W^{\mathrm{sh}})+n_-(W^{\mathrm{sh}})\) holds (with \(n_-(W^{\mathrm{sh}})\) being the number of negative eigenvalues of \(W^{\mathrm{sh}}\)), specifying \(n_+(W^{\mathrm{sh}})\) completely determines \(n_-(W^{\mathrm{sh}})\) and \(\operatorname{sig}{(W^{\mathrm{sh}})}\coloneqq n_+(W^{\mathrm{sh}})-n_-(W^{\mathrm{sh}})\). As a result, the topological invariant can be defined as \(s_1(W)\coloneqq \frac{1}{2}\mathrm{sig}(W^{\mathrm{sh}})\) as in quantum cases~\cite{24photonic}. However, it is more convenient to use the topological invariant
\begin{align}
\begin{split}
	s(W)&\coloneqq n_+(W^{\mathrm{sh}})-1\\
&=n_+(W-E_0\mathbb{I}_n)-1.
\end{split}
\end{align}
This is because \(n_+(W^{\mathrm{sh}})\geq1\) due to the essentially non-negativity constraint Eq.~\eqref{Metzler} and shift Eq.~\eqref{shiftedcon}, making \(s(W)\) take values in \(\mathbb{N}\) as \(n\to\infty\). It thus provides the topological classification for transient states of Hermitian 0D stochastic systems in class AI. Note that this classification derived using \(s(W)\) is equivalent to the previous one in Table~\ref{tabletransientH}. While the phase with \(s=0\) is the trivial phase in the steady-state limit (being the only phase), this is unspecified for transient states.

Besides a unique steady state with eigenvalue \(\lambda_0=0\), all other eigenstates have negative eigenvalues. Hence as \(E_0\to0^-\), the topological invariant becomes \(s(W)=1-1=0\) (see Fig.~\ref{fig:zerodimh}(b)). As \(s_1(W)\to-\infty\) instead, we prefer to use \(s(W)\) to characterize the topology. As any two transition matrices can be continuously deformed into each other in the limit of \(E_0\to0^-\), all Hermitian 0D stochastic systems in class AI are topologically trivial. 

\subsubsection{0D general cases}
\label{sec:0dgeneral}
The result we just obtained that all Hermitian 0D stochastic systems in class AI are trivial in the steady-state limit can be generalized to other symmetry classes, which we will demonstrate by studying \(W\) directly. 

Consider two transition matrices \(W_0\) and \(W_1\) which are gapped at \(E_0\), and denote the eigenvalue of \(W\) with the greatest non-zero real part by \(\lambda_1(W)\). The convex combination 
\begin{align}
W_t=tW_1+(1-t)W_0
\end{align}
of \(W_0\) and \(W_1\) is also a transition matrix. We orthogonally decompose \(\mathbb{R}^n\) into \(\operatorname{span}\{\ket{1}\}\oplus V\) where \(\ket{1}\) is an \(n\)-dimensional column vector of all ones and \(V\) is the corresponding orthogonal complement. Then we can project \(W_t\) onto \(V\) as 
\begin{align}
W_t^{(V)}=(1-t)W_0^{(V)}+tW_1^{(V)}.
\end{align}
Since \(\bra{1}W_t=0\) and the steady state is unique due to the Perron-Frobenius theorem, we have \(\operatorname{spec}(W_t^{(V)})=\operatorname{spec}(W_t)\setminus\{0\}\). Consequently, according to Weyl's inequalities~\cite{12matrix}, we have 
\begin{align}
\lambda_1(W_t^{(V)})\leq(1-t)\lambda_1(W_0^{(V)})+t\lambda_1(W_1^{(V)}).
\end{align}
If the point gap \(E_0\) is chosen such that \(\lambda_1(W_0)<E_0\) and \(\lambda_1(W_1)<E_0\), then \(\lambda_1(W_t^{(V)})<E_0\) holds. 

Therefore, during the deformation, the bulk gap remains open and hence \(W_t\) is a continuous family of transition matrices gapped at \(E_0\) as \(t\) goes from 0 to 1. Thus in these systems, any transition matrix \(W_0\) with \(\lambda_1(W_0)<E_0\) can be continuously deformed into another transition matrix \(W_1\) with \(\lambda_1(W_1)<E_0\) along this simple path. Then in the limit of \(E_0\to 0^-\), \(\lambda_1(W)<E_0\) is automatically satisfied. Therefore, all transition matrices gapped at \(E_0\) can be continuously connected to each other. Since this holds for arbitrary \(n\), Hermitian 0D stochastic systems in class AI are topologically trivial. 

The same argument also applies to stochastic systems in other symmetry classes (including discrete spatial symmetries as well). As \(E_0\to0^{-}\), all phases collapse to the same phase where the spectrum can be divided into two parts. As illustrated in Fig.~\ref{fig:zerodimh}(b), there is a unique steady state and the point gap \(E_0\) in part I, which is the same for all possible cases. In part II, even though the spectrum can be quite complicated, there is no point gap at \(E_0\). Thus any transition matrix can be continuously deformed into any other transition matrix in the same symmetry class since all linear/anti-linear symmetries are preserved along the convex path. Therefore, all Hermitian 0D stochastic systems are topologically trivial. Note that this argument holds only in the steady-state limit. For transient states with a non-zero point gap, Weyl's inequalities are not enough to establish the classification. This explains why we considered the shifted transition matrix \(W^{\mathrm{sh}}\) in Sec.~\ref{sec0daih}. 

Further, this argument holds in non-Hermitian 0D stochastic systems with a line gap, which are effectively Hermitian ones as discussed in Sec.~\ref{pointgap}. More concretely, all non-zero eigenvalues of \(W\) possess a negative real part below \(E_0\)~\cite{12matrix}. Hence as \(E_0\to 0^-\), part I is composed of 0 and a line gap at \(\mathrm{Re}\lambda=E_0\) as in the Hermitian case and thus is the same for all possible cases. Again, although part II can be more complicated with complex eigenvalues, it does not contribute to the classification. Consequently, all non-Hermitian 0D stochastic systems with a line gap are also topologically trivial in the steady-state limit. 

\subsubsection{\((d\geq1)\)-D general cases}
\label{dDHermitian}
The argument that Hermitian stochastic systems are topologically trivial in the steady-state limit remains the same in higher dimensions (\(d\geq1\)). In fact, given a \(\bm{k}\)-space transition matrix \(W(\bm{k})\), ergodicity ensures a unique steady state with eigenvalue \(\lambda_0\equiv0\), analogous to the 0D case. More concretely, we can denote the \(i\)-th eigenvalue of \(W(\bm{k})\) by \(\lambda_i(\bm{k})\). Then it is straightforward to check that \(W(\bm{k}=\bm{0})\) is also a transition matrix with \(\lambda_i(\bm{0})=0\) for a certain \(i\). Since \(\lambda_i(\bm{k})\) is continuous in \(\bm{k}\), we have \(\lambda_i(\bm{k})\to0\) as \(\bm{k}\to\bm{0}\). As the inverse Fourier form \(W\) of \(W(k)\) is a symmetric matrix, \(\lambda_i(\bm{k})\) is a real-valued function and can only approach 0 along the real axis. As \(E_0\to0^-\), \(\lambda_i(\bm{k})\) has to pass through \(E_0\) as \(\bm{k}\to\bm{0}\). Hence, this system is gapless in the steady-state limit.

Similar to the generalization in the last section, this argument further holds in non-Hermitian stochastic systems with a line gap as well. Even though \(\lambda_i(\bm{k})\) is no longer a real-valued function, \(\lambda_i(\bm{k})\) with \(\bm{k}\neq0\) is non-zero and thus has a negative real part. Since the real part \(\mathrm{Re}\lambda_i(\bm{k})\) is also continuous, as \(E_0\to0^-\), it has to pass through the line gap \(\mathrm{Re}\lambda=E_0\). Hence, this system is again gapless in the steady-state limit.

Thus, our work can explain why all Hermitian stochastic systems are topologically trivial in the steady-state limit, consistent with previous work~\cite{24nonre}. Hence, to predict non-trivial topological phases in the steady state, we have to consider non-Hermitian stochastic systems, as will be shown in the following sections.


\begin{figure}[htbp]
\centering
\includegraphics[scale=0.3]{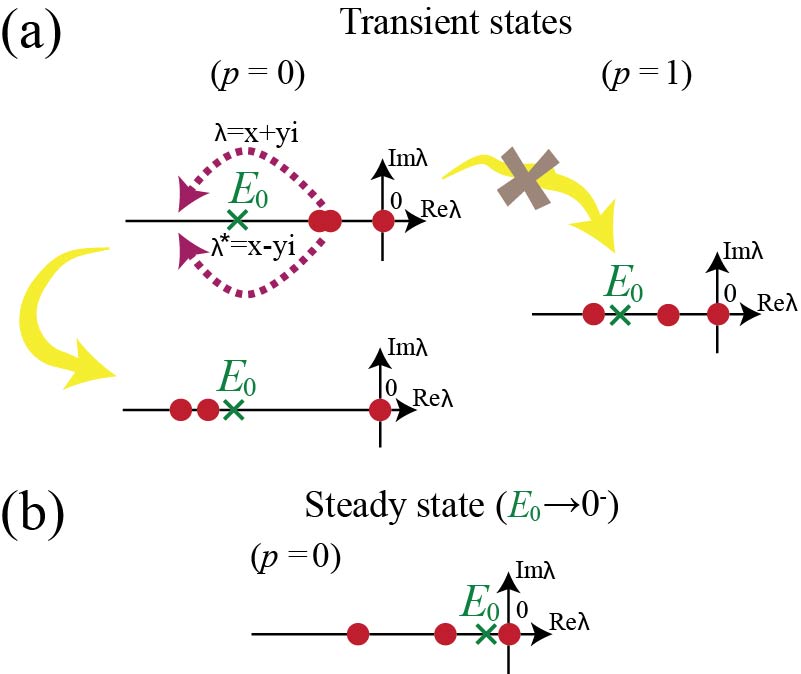}
\caption{Restrictions on the motion of eigenvalues \(\lambda\) of \(W\) (red dots) in non-Hermitian 0D stochastic systems in class AI for (a) transient states and (b) the steady state. The topological invariant is defined as \(p\coloneqq n_-\bmod2\), the parity of the number of real eigenvalues below the point gap \(E_0\) (green cross). (a) With \(E_0\neq0\), the phases with \(p\)=0 can be continuously deformed into each other (yellow arrow). However, they cannot be continuously deformed into the phase with \(p\)=1 (blocked arrow) because the real nature of \(W\) restricts each eigenvalue to appear as either a single real number or one of a complex conjugate pair \(\lambda=x+y\mathrm{i}\) and \(\lambda^*=x-y\mathrm{i}\). (b) In the steady-state limit (\(E_0\to0^-\)), only the phase with \(p=0\) remains and hence must be the trivial phase.}
\label{fig:zerodimnh}
\end{figure}

\subsection{Non-Hermitian stochastic systems}
\subsubsection{0D class AI}
\label{sec:0dainonh}
In non-Hermitian systems, to construct the topological invariant for \(W^{\mathrm{sh}}\), we can also use an eigenvalue-tracking method similar to the previous Hermitian case in Sec.~\ref{sec0daih}. A complex eigenvalue \(\lambda\) of \(W^{\mathrm{sh}}\in\mathcal{W}_\textrm{AI}(n,-E_0)\) generally possesses a counterpart \(\lambda^*\) in the spectrum since \(W^{\mathrm{sh}}\) is real. This can be illustrated in the original space of \(W\), see Fig.~\ref{fig:zerodimnh}(a). The complex conjugate pair \(\lambda\) and \(\lambda^*\) can be moved around the point gap without passing through it (indicated by the dashed arrows), whereas a single real eigenvalue cannot pass through the point gap. Hence, \(W\) with an even number of eigenvalues below \(E_0\) cannot be continuously deformed into another \(W\) with an odd number of eigenvalues below \(E_0\). 

Since any pair of negative eigenvalues can be merged into such a complex pair and deformed, only the parity of \(n_-(W^{\mathrm{sh}})\equiv n_-(W-E_0\mathbb{I}_n)\) is well-defined and characterizes the topology. Thus, the topological invariant can be defined as
\begin{align}
\begin{split}
p(W)&\coloneqq n_-(W^{\mathrm{sh}})\bmod2\\
&=n_-(W-E_0\mathbb{I}_n)\bmod2.
\end{split}
\end{align}

This \(\mathbb{Z}_2\) invariant characterizes the topological phases for transient states of 0D stochastic systems in class AI. In fact, since \(\det{W^{\mathrm{sh}}}=\prod^{n-1}_{i=0}\lambda_i(W^{\mathrm{sh}})\), we have \((-1)^{p(W)}=\operatorname{sgn}\det{(W-E_0\mathbb{I}_n)}\), just as in the quantum case~\cite{24photonic}. To verify this, it is straightforward to check that matrices with different signs of the determinant belong to different phases since the determinant is a continuous function of the matrix. Conversely, shifted transition matrices \(W^{\mathrm{sh}}\) with the same parity can be continuously deformed into one another. Note that in a discrete-time classical stochastic process, the parity \(p\) determines whether a stochastic matrix can be realized with any Markovian master equation~\cite{26discrete}.

In the limit of \(E_0\to0^-\), the situation is similar to the Hermitian case except for certain complex conjugate pairs in the spectrum. However, since \(p(W)\) depends only on the parity of the real eigenvalues and is not affected by complex pairs, only one phase remains, characterized by \(p(W)=(n-1) \bmod 2\), as illustrated in Fig.~\ref{fig:zerodimnh}(b). Therefore, the steady state in 0D stochastic systems in class AI is also topologically trivial.

\subsubsection{1D class AI}
\label{onedAI}
We now consider one-dimensional (1D) stochastic systems in class AI. We showed in Sec.~\ref{subsec:homotopy_steady}, that the topological classification of stochastic systems with finite \(E_0\) (corresponding to transient states) is the same as in the quantum case and as a result, we can use the same real-space invariant and \(\bm{k}\)-space invariant. More concretely, the \(\bm{k}\)-space topological invariant for the 1D transition matrix \(W(k)\) is the conventional winding number~\cite{17stochastic,24role}
\begin{align}
\nu\coloneqq \int_0^{2\pi}\frac{dk}{2\pi\mathrm{i}}\partial_k\log\det{(W(k)-E_0\mathbb{I}_n)},
\label{kwind}
\end{align}
analogous to the quantum one~\cite{18edge,19symm}. The explicit construction of the real-space homotopy invariant and its equivalence with the \(\bm{k}\)-space invariant are provided in Appendix~\ref{sec:onedwind}. This winding number \(\nu\) distinguishes the topological phase from the trivial one~\cite{11TI} and supports a protected boundary mode~\cite{24ergodic}.

\begin{figure}[htbp]
\centering
\includegraphics[width=0.48\textwidth]{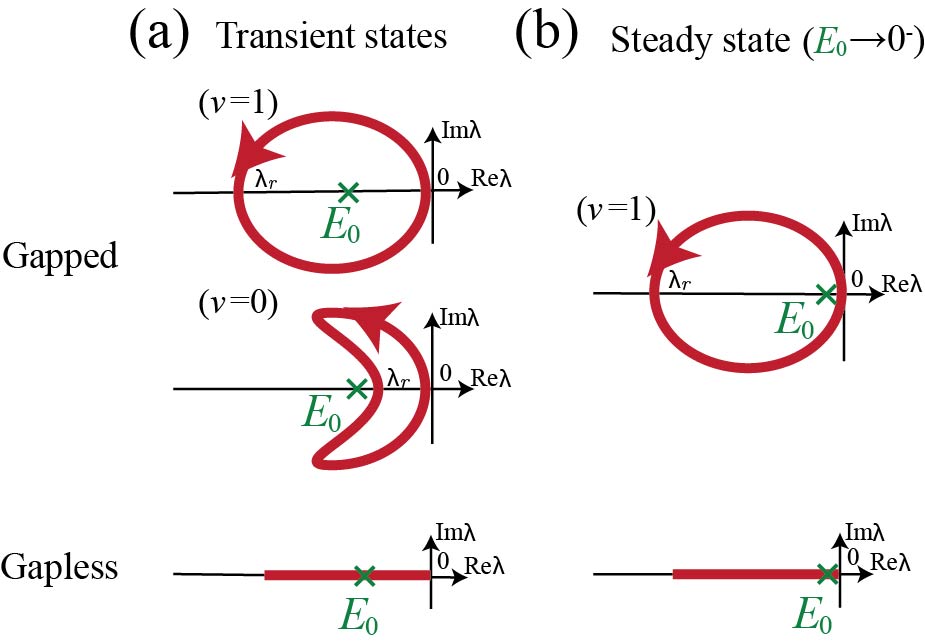}
\caption{PBC spectra of non-Hermitian 1D stochastic systems in class AI for (a) transient states and (b) the steady state. The topological invariant is the winding number \(\nu\) of the red loops composed of eigenvalues \(\lambda\) of \(W(k)\) with respect to a reference point \(E_0\) (green cross). However, gapless spectra (red line segments, below) include \(E_0\) and hence \(\nu\) is ill-defined. (a) Two phases of gapped cases and a gapless case are shown. Loops with \(\nu=1\) and \(\lambda_r<E_0\) are topologically different from loops with \(\nu=0\) and \(\lambda_r>E_0\) where \(\lambda_r\) is a non-zero real eigenvalue of \(W(k)\). (b) In the steady-state limit (\(E_0\to0^-\)), \(\lambda_r<E_0\) always holds. Hence phases with gapped spectra will always collapse to the phase with \(\nu\neq 0\) while the gapless spectrum will remain gapless. As a result, phases in 1D class AI with gapped spectra are always non-trivial while phases with gapless spectra are always trivial.}
\label{fig:onedim}
\end{figure}

For transient states, as illustrated in Fig.~\ref{fig:onedim}(a), this winding number \(\nu\) simply counts the number of times that a loop (of red dots) representing the spectrum of \(W(k)\) winds around the reference point \(E_0\) (green cross). As stated in Sec.~\ref{sec:markovian}, the topological invariant is only well-defined for gapped systems represented by loops but cannot be defined in gapless systems where \(E_0\) coincides with an eigenvalue in the spectrum. This explains why in Fig.~\ref{fig:onedim}(a), both loops representing gapped systems support a well-defined \(\nu\).

To characterize different phases, we examine the limit \(E_0\to0^-\). As \(k\) goes from 0 to \(2\pi\), the eigenvalues pass through \(\lambda_0=0\) once due to the Perron-Frobenius theorem. Hence, as illustrated in Fig.~\ref{fig:onedim}(b), loops with a non-zero winding number can exist, while loops passing through \(\lambda_0=0\) with \(\nu=0\) cannot. In fact, a non-zero real eigenvalue of \(W(k)\) can be denoted as \(\lambda_r\). Then \(\lambda_r<E_0\) always holds as \(E_0\to0^-\). Thus it is impossible to construct a loop with \(\nu=0\) because of the continuity of \(\lambda(k)\) (with details provided in Appendix~\ref{sec:nozeroloop}). Consequently, the winding numbers are fixed at \(|\nu|=1\) in phases with gapped spectra, which are topological phases. 

Conversely, gapless systems are trivial phases since in this case it is impossible to construct a well-defined winding number. As illustrated in the bottom part of Fig.~\ref{fig:onedim}, if the spectrum extends from 0 along the real axis (indicated by a line segment), the system is gapless as defined in Sec.~\ref{pointgap}. Since a gapless spectrum remains gapless even in the steady-state limit, it corresponds to the trivial phase.

In summary, for 1D stochastic systems in class AI, the phase with a gapped spectrum is always non-trivial and the phase with a gapless spectrum is always trivial, in the steady-state limit. The topological classification is thus given by \(\nu\in\mathcal{Z}_2\coloneqq\{-1,1\}\). Note that \(\mathcal{Z}_2\) is just a set rather than the additive group \(\mathbb{Z}_2\), since addition is not permitted for ergodic systems that have a unique steady state. Once we consider multiple systems such as by stacking them~\cite{22boundary}, addition is permitted.

\subsubsection{\((d\geq2)\)-D class AI}
In higher dimensions, the topological classification for transient states of stochastic systems in class AI can be derived from the results for their quantum counterparts in class AI~\cite{19symm} as in the 1D case. For \(d\geq2\), non-trivial cases only occur in \(d=5\) and \(d=7\). Hence we focus on the classification for the steady state in these cases.

In 5D, the topological invariant \(\nu_5\) is the 5D winding number~\cite{10dimensional,16classification}. Unlike the integer-valued 1D winding number \(\nu\), \(\nu_5\) takes values in \(2\mathbb{Z}\) as in quantum cases. This even winding number \(\nu_5\) might appear to be forbidden by the uniqueness of the steady state, but we will see that this is not the case. In 1D, \(\nu\) corresponding to \(\pi_1(U(1))\cong\pi_1(S^1)\) is a spectral winding number characterized by the phase of \(\operatorname{det}{W^{\mathrm{sh}}(k)}\), which encodes the winding information of the eigenvalues with respect to \(E_0\). In contrast, because of the additional dimensions in 5D, \(\nu_5\) corresponding to \(\pi_5(\operatorname{U}(n))\cong\pi_5(\operatorname{SU}(n))\) can be obtained from \(\bm{k}\mapsto W^{\mathrm{sh}}(\bm{k})\in \operatorname{SU}(n)\) with large enough \(n\). Since any matrix \(W^{\mathrm{sh}}(\bm{k})\) in \(SU(n)\) must satisfy \(\det W^{\mathrm{sh}}(\bm{k})=1\), \(\nu_5\) is insensitive to \(\operatorname{det}{W^{\mathrm{sh}}(\bm{k})}\) and the spectral winding behavior. As a result, an even-valued \(\nu_5\) is compatible with ergodicity and a unique steady state. Consequently, ergodicity can only modify the classification results in low dimensions. Therefore, the topological classification for the steady state of 5D stochastic systems in class AI is also \(2\mathbb{Z}\).

On the other hand, the topological invariant \(\nu_7\) in 7D is defined as the parity of the 7D Chern-Simons invariant \(CS_7\)~\cite{10dimensional,16classification}. As \(E_0\to0^-\), analogous to the 5D case, the dimension is high enough to allow \(CS_7\) to take all possible values. Hence the topological classification for the steady state of 7D stochastic systems in class AI is given by \(\mathbb{Z}_2\). Accordingly, the full topological classification of class AI stochastic systems for the steady state is summarized in Table~\ref{table:nonH}.

\begin{table}[htbp]
\centering
\small
\setlength{\tabcolsep}{4pt}
\caption{Topological classification of the steady state of all stochastic systems. We show that only non-Hermitian systems can support topologically non-trivial phases. Only class AI with no symmetry survives the constraints on a transition matrix, enumerated by spatial dimension \(d\). Since the period of dimensions is 8 as in quantum systems, only dimensions up to 7D are listed. Entries give the group of topological invariants available in each case: \(0\) where only the trivial phase exists, and \(\mathbb{Z}_2\), or \(2\mathbb{Z}\) otherwise. Note that \(\mathcal{Z}_2\) denotes the doubleton \(\{-1,1\}\) without group structure.}
\label{table:nonH}
\begin{tabular}{|c|c|c|c|c|c|c|c|c|}
\hline\hline
\diagbox[height=1.5em]{Class}{\quad\(d\)}&\;0\; &\;1\; &\;2\; &\;3\; &\;4\; &\;5\; &\;6\; &\;7\;\\
\hline\hline
\multicolumn{9}{|c|}{\textbf{Non-Hermitian systems}}\\
\hline
AI & 0 & \(\mathcal{Z}_2\) & 0 & 0 & 0 & \(2\mathbb{Z}\) & 0 & \(\mathbb{Z}_2\)\\
\hline
\end{tabular}
\end{table}

\subsubsection{\(d\)-D class BDI\(^\dagger\)}
\label{sec:BDIdagger}
As discussed in Sec.~\ref{sec:classtransient}, the classifications are the same for non-Hermitian class BDI\(^\dagger\) and Hermitian class AI when \(E_0\) is finite. In fact, the corresponding classification in the limit of \(E_0\to0^-\) is also the same. More concretely, even though a shifted transition matrix \(W^\mathrm{sh}\) in class BDI\(^\dagger\) is non-Hermitian, it suffices to study the Hermitian matrix \(D=WC_+\) in the steady-state limit. As a result, we can apply the same method to class BDI\(^\dagger\) as for the Hermitian case of class AI discussed in Secs.~\ref{sec:0dgeneral} and \ref{dDHermitian}.

Therefore, in the steady-state limit, stochastic systems in class BDI\(^\dagger\) are also trivial. This leaves only one symmetry class that hosts robust steady states in stochastic systems as shown in Table~\ref{table:nonH}, reflecting the full topological classification of steady states.

\section{Conclusion}
The gapless nature of general transition matrices leads to a fundamental obstacle to the establishment of a well-defined topological classification in stochastic systems. To address this issue, we define a point gap for transition matrices, thereby yielding properly gapped systems. Our definition provides the first unified framework for all discrete non-spatial symmetries and dimensions, enabling their topological classification. Since standard \(K\)-theoretic arguments used in quantum systems are not compatible with Markovian constraints, we instead use a homotopy approach to establish the topological classification of stochastic systems. 

Specifically, we prove the homotopy equivalence between the space of both point-gapped and line-gapped transition matrices and the standard classifying space well studied in the quantum case. Hence, our approach establishes the topological correspondence between quantum and stochastic systems thus explains why one can directly use the topological invariants defined in quantum cases to characterize the topology of stochastic systems~\cite{17stochastic,18dissipative,24role}. Moreover, as explained in Sec.~\ref{sec:homotopyequiv}, we have established the homotopy equivalence between spaces of transition matrices, non-negative matrices, stochastic matrices, and doubly stochastic matrices~\cite{12matrix}. Hence, our homotopy approach can also be applied to discrete-time stochastic process~\cite{26discrete} characterized by stochastic matrices. The methods we develop are also relevant for similar issues in the topological classification of quantum Lindbladians~\cite{23symm}, and thus open new directions for many other systems. 

We find that only the symmetry classes AI and BDI\(^\dagger\), which are no symmetry and pseudo-Hermiticity respectively, support non-trivial topology. Other symmetries are either topologically unstable against perturbations, or forbidden by the Markovian constraints. Based on whether the point gap is finite or approaching 0, we classify both transient states (Table~\ref{tabletransientH}) and steady states (Table~\ref{table:nonH}). For transient states, we prove that the classification inherits the quantum one~\cite{19symm} for class AI, which is the only class with non-trivial transient states in Hermitian stochastic systems. For the steady state, the classification is new due to having to take the limit of the point gap. 

We find topologically protected steady states only in the non-Hermitian class AI, which encompasses generic cases such as random networks, and is consistent with many realistic stochastic systems showing little evidence of symmetry~\cite{20fibration,22quasifibrations}. 
In low dimensions (\(d\leq3\)), only 1D systems in class AI support topological phases characterized by a \((\pm1)\)-valued winding number with gapped (gapless) spectra being always topologically non-trivial (trivial), while higher dimensions can host integer-valued topological invariants. In addition, we examine non-Hermitian stochastic systems with a line gap to show that their classification is the same as their Hermitian counterpart, just like in the quantum case~\cite{19symm}. These results explain in a rigorous way why all Hermitian ergodic stochastic systems are topologically trivial in the steady state and hence, non-Hermiticity is necessary for the existence of non-trivial topology, a key difference between quantum and stochastic systems first proposed by observing the steady-state distribution~\cite{24nonre}. 

This work forms a basis for several other future lines of work. Even though we focus on the typical ergodic case, our homotopy approach to establish the topological classification and construct the topological invariants does not depend on ergodicity. Hence, it is straightforward to generalize our approach to non-ergodic stochastic systems, such as simply by stacking ergodic systems~\cite{22boundary}. In addition, the effect of exceptional points on classification results is an avenue for further study, especially with the use of new methods such as Morse theory~\cite{26morse}. 
Further, as our work addresses discrete non-spatial symmetries, it would be of interest to generalize this framework to other spatial symmetries such as crystalline ones~\cite{21topology, 24mechanism, 26morse}. Another open question is the topological classification of stochastic systems with separable eigenvalue bands (where separability is defined by eigenvalues satisfying \(\lambda_n(\bm{k}) \neq \lambda_m(\bm{k})\) for all \(m \neq n\) and all \(\bm{k}\))~\cite{13braid,18band,20homotopy}. Lastly, the real-space invariants we have developed to characterize stochastic systems are applicable to systems without translation invariance~\cite{15ktheory, 19real}, such as in disordered systems and amorphous networks~\cite{21design, 22ambi,25realspace}. Such directions show the many avenues of interest opened by this new classification of topological phases of matter, with implications for identifying robust steady states and observables in biochemical and synthetic systems. 

\begin{acknowledgments}
We thank Alexander Cerjan for helpful discussions on the spectral localizer, and also Ryan Budney and Thomas Goodwillie for their useful mathematical insights. This work has been supported by the NSF Center for Theoretical Biological Physics (PHY2019745), the NSF CAREER Award (DMR-2238667), and the CZI Theory Institute Without Walls.
\end{acknowledgments}

\appendix

\section{Detailed explanation of issues in stochastic systems}
\subsection{Why does the spectral flattening technique fail in stochastic systems?}
\label{sec:failflat}
In this section, we explain why the spectral flattening is not feasible in the stochastic systems. Since the spectral flattening of non-Hermitian matrices can also be reduced to Hermitian matrices~\cite{19symm}, we can focus on Hermitian cases.

In quantum systems, given a symmetric Hamiltonian \(H\), we can diagonalize it as \(H=OD O^T\), where \(D\) is a diagonal matrix composed of the spectrum of \(H\). Then it is possible to construct a continuous path of diagonal matrices \(D(t)\) to deform \(D(0)=D\) into \(D(1)\) satisfying \(D(1)^2=\mathbb{I}_n\)~\cite{20onsite,24homotopy}. During the continuous deformation, \(H(t)=OD(t) O^T\) is a matrix in the original symmetry class and is invertible if and only if \(D(t)\) is invertible. Consequently, we have \(H(1)^2=OD(1)O^TOD(1)O^T=\mathbb{I}_n\). Since the spectrum of \(H(1)\) is fixed, only the orthogonal matrix \(O\in O(n)\) determines the topology~\cite{16classification}.

Inspired by the technique in quantum systems, given a symmetric shifted transition matrix \(W^{\mathrm{sh}}\) in stochastic systems, we can apply the spectral decomposition and obtain the decomposition \(W^{\mathrm{sh}}=OD'O^T\) with the orthogonal matrix \(O\) as the transformation matrix and the diagonal matrix \(D'\) as the spectral matrix. Hence we can also construct the corresponding one-parameter family of matrices
\begin{align}
\{W^{\mathrm{sh}}(t)=OD'(t)O^T\mid t\in[0,1]\}.
\end{align}
Our goal is to flatten the spectral matrix \(D'(t)\) and thus flatten the spectrum of \(W^{\mathrm{sh}}(t)\) at \(t=1\). However, at each \(t\), \(W^{\mathrm{sh}}(t)\) doesn't have to satisfy shifted Markovian constraints Eqs.~\eqref{Metzler} and \eqref{shiftedcon} even if the invertibility of \(D'(t)\) is ensured. More concretely, in the quantum case, the transformation matrix \(O\) is an orthogonal matrix and thus a real matrix due to the spectral theorem~\cite{12matrix}. Hence \(H(t)\) is real if and only if the spectral matrix \(D(t)\) is real. However, since \(O\) always contains negative entries, \(W^{\mathrm{sh}}(t)\) doesn't satisfy the essentially non-negativity constraint Eq.~\eqref{Metzler} unless the diagonal matrix \(D'(t)\) is deformed carefully, which is generally difficult to ensure for arbitrary \(W^{\mathrm{sh}}\). Hence it is difficult to connect any gapped shifted transition matrix to a flattened transition matrix. 

\subsection{Mismatch of the symmetry between \(H\) and \texorpdfstring{\(H(\bm{k})\)}{Lg} in quantum systems}
\label{sec:mismatch}
As discussed in the main text, in stochastic systems, the Markovian constraints Eqs.~\eqref{Metzler} and \eqref{probcon} are satisfied by \(W\) rather than \(W(\bm{k})\). Hence it is difficult to check whether \(W(\bm{k})\) is the Fourier form of a transition matrix. In this section, we will show how this can be addressed in quantum systems by the method of equivariant homotopy~\cite{14bloch}.

For instance, let us consider quantum systems in class AI. The time-reversal symmetry allows one to choose a basis in which \(H\) is a real matrix, while the \(\bm{k}\)-space Hamiltonian \(H(\bm{k})\) is an invertible complex matrix in the general linear group \(\mathrm{GL}(n,\mathbb{C})\). Nevertheless, the time-reversal symmetry induces an involution \(\tau_1:\bm{k}\mapsto-\bm{k}\) on \(S^d\) and also applies naturally to \(H(\bm{k})\) as an involution \(\tau_2\) gluing \(H(\bm{k})\) and \(H^\ast(-\bm{k})\) together. Hence the space of \(H(\bm{k})\) can be regarded as the equivariant mapping space Map(\((S^d,\tau_1),(\mathrm{GL}(n,\mathbb{C}),\tau_2)\)) consisting of the map \(\bm{k}\mapsto H(\bm{k})\). The corresponding \(\mathbb{Z}_2\)-equivariant homotopy group \(\pi^{\mathbb{Z}_2}_d(\mathrm{GL}(n,\mathbb{C}))\) is just the usual \(K\)-group. Note that here \(\bm{k}\) is taken from \(S^d\) rather than the Brillouin zone \(T^d\) to obtain the strong topological invariants~\cite{09Kitaev}.

However, this method is not applicable to stochastic systems. More concretely, in quantum systems, the time-reversal symmetry generates a \(\mathbb{Z}_2\) group action, which explains why we can consider the equivariant homotopy~\cite{14bloch}. However, in stochastic systems, the essentially non-negativity constraint Eq.~\eqref{Metzler} on \(W\) is not naturally ensured by a symmetry associated with a group. Therefore, the equivariant homotopy approach used above does not directly apply.

\section{Why are the symmetries \(C_-\) and \(S\) not topologically robust?}
\label{sec:non-robust}
We first rewrite Eq.~\eqref{subl} as
\begin{align}
SWS^{-1}=-W+2\frac{\mathrm{Tr}W}{n}\mathbb{I}_n.
\end{align}
Here, the extra term is due to the minus sign in Eq.~\eqref{subl}. We can verify the relation for the diagonal entries
\begin{align}
(SWS^{-1})_{ii}=-W_{ii}+2\frac{\mathrm{Tr}W}{n}.
\label{sublw}
\end{align}
To be more specific, we assume \(S=\mathbb{I}_p\oplus(-\mathbb{I}_q)\) with \(p+q=n\). Then the diagonal entries satisfy
\begin{align}
W_{ii}=\frac{\mathrm{Tr}W}{n}.
\label{diagcons}
\end{align}
Combined with the probability conservation, Eq.~\eqref{diagcons} requires that the sums of the off-diagonal entries in each column of \(W\) are all equal to the same constant. 

However, topological properties have to be insensitive to symmetry-preserving random perturbations~\cite{13intro} while the off-diagonal column sums of the perturbation have to be fixed to protect the topology. Even though a specific transition matrix \(W\) satisfying Eq.~\eqref{sublw} might be found, the perturbations generally break the symmetry if the off-diagonal column sums are not fixed. As a result, \(S=\mathbb{I}_p\oplus(-\mathbb{I}_q)\) cannot support a topologically non-trivial phase against perturbations. For \(C_-\) with the anticommutation relation, the constraint on perturbations due to the extra identity term is even more complicated and the corresponding phase is also not robust against perturbations. Therefore, both \(C_-\) and \(S\) are not considered for the topological classification in the main text.

Moreover, because of probability conservation Eq.~\eqref{probcon}, the open boundary conditions (OBC) in stochastic systems differ from the quantum ones in their diagonal terms~\cite{24role}. Hence if a stochastic system under the periodic boundary conditions (PBC) preserves \(C_-\) or \(S\) with a minus sign in the anticommutation relation, this system under OBC does not preserve these symmetries. As a result, the bulk-boundary correspondence is broken in stochastic systems in these classes. Moreover, such a breakdown is due to the mismatch of classes imposed by the Markovian constraints rather than the non-Hermitian skin effect~\cite{18edge} and thus cannot be addressed by introducing the generalized-Brillouin-zone (GBZ) approach. Note that in quantum systems, however, since there is no need to modify the OBC to satisfy probability conservation Eq.~\eqref{probcon}, these symmetries are proven to protect topological phases~\cite{22shifted}. 

\section{Proof of the homotopy equivalences between transition matrices and real matrices}
\label{sec:homotopyequiv}
\subsection{A unified framework for proving homotopy equivalences}
\label{sec:unified}
In this section, we will propose a unified framework for the proof of those homotopy equivalences in both Hermitian and non-Hermitian systems. Before we study the homotopy, let's first introduce a more convenient space \(\mathcal{W}(n,1)\). More concretely, by continuously rescaling all matrices in \(\mathcal{W}(n,-E_0)\) by the factor \(\frac{1}{-E_0}\), we show that \(\mathcal{W}(n,-E_0)\) is homeomorphic to \(\mathcal{W}(n,1)\). Hence in \(\mathcal{W}(n,1)\), probability conservation Eq.~\eqref{shiftedcon} is now normalized as
\begin{align}
\sum^n_{i=1}W^{\mathrm{sh}}_{ij}=1,\forall j.
\label{shiftedcon1}
\end{align}

We first show that the normalized probability conservation Eq.~\eqref{shiftedcon1} reduces the effective order of \(W^\mathrm{sh}\) by 1. Define the isotropy group \(\mathrm{Aff}_{1}(n-1,\mathbb{R})\) of \(\mathrm{GL}(n,\mathbb{R})\) satisfying the normalized probability conservation Eq.~\eqref{shiftedcon1}. First, we have to prove \(\mathrm{Aff}_{1}(n-1,\mathbb{R})\simeq\mathrm{GL}(n-1,\mathbb{R})\). Define an affine group \(\mathrm{Aff}(n-1,\mathbb{R})\) as the isotropy group of \(\mathrm{GL}(n,\mathbb{R})\) leaving \(\bra{e_1}\) invariant with \(\bra{e_1}\coloneqq(1,0,\dots,0)\) the first standard basis vector of \(\mathbb{R}^{n*}\). Since there exists an invertible matrix \(O\) satisfying \(\bra{e_1}O=\frac{1}{\sqrt{n}}\bra{1}\), given \(G\in\mathrm{Aff}_{1}(n-1,\mathbb{R})\), we have \(\bra{e_1}OGO^{-1}=\frac{1}{\sqrt{n}}\bra{1}GO^{-1}=\frac{1}{\sqrt{n}}\bra{1}O^{-1}=\bra{e_1}\). Hence we have \(OGO^{-1}=\begin{pmatrix}
1&\bra{0}\\\ket{v}&C
\end{pmatrix}\) in the basis of \(\{\ket{e_i}\}_i\). Here, \(C\) is an \((n-1)\times(n-1)\) real matrix while \(\ket{v}\in\mathbb{R}^n\) and \(\bra{0}\in\mathbb{R}^{n*}\) are an arbitrary vector and a null vector respectively. Then the point-gapped condition is written as \(\det G=\det C\neq0\) and hence all possible \(C\)'s form a space \(\mathrm{GL}(n-1,\mathbb{R})\). Therefore, \(\mathrm{Aff}_1(n-1,\mathbb{R})\) is homotopy equivalent to \(\mathrm{GL}(n-1,\mathbb{R})\). 

Then we show that the essentially non-negativity constraint Eq.~\eqref{Metzler} can be removed from a homotopical perspective. It suffices to prove \(\mathcal{W}(n,1)\simeq\mathrm{Aff}_{1}(n-1,\mathbb{R})\). More concretely, we can define a map \(H\) on \(\mathrm{Aff}_{1}(n-1,\mathbb{R})\times [0,1]\) by 
\begin{align}
H(G, t)\coloneqq \frac{G+t\mu(G)\ket{1}\bra{1}}{1+nt\mu(G)}
\end{align}
with \(\mu(G)=\max\{0,-\min_{i\neq j}G_{ij}\}\geq0\). We have 
\begin{align}
\begin{split}
&OH(G, t)O^{-1}\\
=&O\frac{G+t\mu(G)\ket{1}\bra{1}}{1+nt\mu(G)}O^{-1}\\
=&\frac{1}{1+nt\mu(G)}(\begin{pmatrix}
1&\bra{0}\\\ket{v}&C
\end{pmatrix}+\begin{pmatrix}
nt\mu(G)&\bra{0}\\\ket{0}&\textbf{0}_{n-1}
\end{pmatrix})\\
=&\frac{1}{1+nt\mu(G)}\begin{pmatrix}
1+nt\mu(G)&\bra{0}\\\ket{v}&C
\end{pmatrix}
\end{split}
\end{align}
with \(\bm{0}_{n-1}\) the \((n-1)\times(n-1)\) null matrix. Since \(1+nt\mu(G)\geq1\), \(H(G, t)\) is invertible if and only if \(G\) is invertible. Then it is straightforward to prove that \(\mathcal{W}(n,1)\) is a strong deformation retract of \(\mathrm{Aff}_{1}(n-1,\mathbb{R})\).

Note that the method in this section can be directly applied even to stochastic systems with a line gap, hence unifying our classification framework for both point-gapped and line-gapped scenarios. Moreover, this also works for Hermitian cases by simply requiring \(\ket{v}\) to be a null vector. Therefore, this unified framework can already work in all cases considered in this work. It is thus unnecessary to introduce non-negative matrices and stochastic matrices if we are just interested in the proof of homotopy equivalences. However, as we will discuss in Secs.~\ref{sec:classtransient} and \ref{sec:onedwind}, non-negative matrices are formally much simpler with fewer constraints and can be used to construct the topological invariants. Moreover, stochastic matrices also play an important role as stated in Sec.~\ref{sec:non-negative}. As a result, we will show how to prove the homotopy equivalences via non-negative and stochastic matrices for both non-Hermitian and Hermitian stochastic systems in the following sections.

\subsection{Non-Hermitian case}
\label{sec:homotopynonh}
\subsubsection{\(\mathcal{W}(n,-E_0)\simeq\mathcal{A}(n)\)}
Our first goal is to prove that \(\mathcal{W}(n,1)\) and \(\mathcal{A}(n)\coloneqq\{A\in\mathrm{GL}(n,\mathbb{R})|A_{ij}\geq0\}\), the space of all invertible entrywise non-negative matrices, have the same homotopy type. Up to homotopy, this means that shifted Markovian constraints Eqs.~\eqref{Metzler} and \eqref{shiftedcon1} on \(W^{\mathrm{sh}}\) can be replaced by the non-negativity constraint Eq.~\eqref{non-negativity} as in the main text.

To prove \(\mathcal{W}(n,1)\simeq\mathcal{A}(n)\), we construct a space \(\mathcal{P}(n)\coloneqq\{P\in\mathcal{A}(n)\mid\bra{1}P=\bra{1}\}\) of \(n\times n\) invertible stochastic matrices~\cite{12matrix}. We first prove that \(\mathcal{P}(n)\) is a strong deformation retract of \(\mathcal{A}(n)\). Given any \(A\in\mathcal{A}(n)\), we define a diagonal matrix \(D_{\mathrm{col}}\) by 
\begin{align}
	D_{\mathrm{col}}\coloneqq\mathrm{diag}(1/d_1,\dots,1/d_n)
\end{align}
with \(d_j=\sum^n_{i=1}A_{ij}>0\) since \(A\) is non-negative and invertible. Hence the equality
\begin{align}
\alpha(A)\coloneqq AD_{\mathrm{col}}
\label{fcol}
\end{align}
defines a map \(\alpha:\mathcal{A}(n)\to\mathcal{P}(n)\) that scales each column sum to \(1\). Then we define a map \(H_1\) on \(\mathcal{A}(n)\times[0,1]\) by
\begin{align}
H_1(A, t)\coloneqq AD_{\mathrm{col}}(t),
\end{align}
where \(D_{\mathrm{col}}(t)\) is a family of diagonal matrices defined by
\begin{align}
D_{\mathrm{col}}(t)\coloneqq (1-t)\mathbb{I}_n+tD_{\mathrm{col}}.
\end{align}
Since \(D_{\mathrm{col}}(t)\in\mathcal{A}(n)\) is invertible and non-negative, we have \(H_1(A, t)\in\mathcal{A}(n)\). Hence the map \(H_1\) is actually a homotopy. Moreover, it is straightforward to verify that \(H_1(A,0)=A\) and \(H_1(A,1)=\alpha(A)\in\mathcal{P}(n)\). Lastly, for any \(P\in\mathcal{P}(n)\), we have \(D_{\mathrm{col}}=\mathbb{I}_n\) and thus \(H_1(P, t)=P\). Thus \(\mathcal{P}(n)\) is a strong deformation retract of \(\mathcal{A}(n)\).

It is thus enough to prove \(\mathcal{W}(n,1)\simeq\mathcal{P}(n)\). We can define a map \(H_2\) on \(\mathcal{W}(n,1)\times [0,1]\) by 
\begin{align}
H_2(W^{\mathrm{sh}}, t)\coloneqq \frac{W^{\mathrm{sh}}+t\mu_2(W^{\mathrm{sh}})\ket{1}\bra{1}}{1+nt\mu_2(W^{\mathrm{sh}})}
\end{align}
with \(\mu_2(W^{\mathrm{sh}})=\max\{0,-\min_{i, j}W^{\mathrm{sh}}_{ij}\}=\max\{0,-\min_{i}W^{\mathrm{sh}}_{ii}\}\). It is straightforward to check \(H_2(W^{\mathrm{sh}}, t)\in\mathcal{W}(n,1)\) at each \(t\) as in Sec.~\ref{sec:unified}. Then it suffices to prove that \(H_2\) is a retraction from \(\mathcal{W}(n,1)\) onto \(\mathcal{P}(n)\). It is straightforward to check \(H_2(W^{\mathrm{sh}},0)=W^{\mathrm{sh}}\). Moreover, given \(A\in\mathcal{P}(n)\), we have \(\mu_2(A)=0\) and \(H_2(A, t)=A\). Finally, the \(i\)-th diagonal entry \((H_2(W^{\mathrm{sh}},1))_{ii}=\frac{W^{\mathrm{sh}}_{ii}+\mu_2(W^{\mathrm{sh}})}{1+n\mu_2(W^{\mathrm{sh}})}\) is non-negative due to \(W^{\mathrm{sh}}_{ii}+\mu_2(W^{\mathrm{sh}})=W^{\mathrm{sh}}_{ii}\) if \(W^{\mathrm{sh}}\) is non-negative and \(W^{\mathrm{sh}}_{ii}+\mu_2(W^{\mathrm{sh}})\geq0\) otherwise. This means \(H_2(W^{\mathrm{sh}},1)\in\mathcal{P}(n)\). Therefore, \(\mathcal{P}(n)\) is a strong deformation retract of \(\mathcal{W}(n,1)\) and we thus have \(\mathcal{W}(n,1)\simeq\mathcal{P}(n)\simeq\mathcal{A}(n)\).

\subsubsection{\(\mathcal{A}(n)\simeq\mathrm{GL}(n-1,\mathbb{R})\)}
Intuitive proofs for the homotopy equivalence between \(\mathcal{A}(n)\) and \(\mathrm{GL}(n-1,\mathbb{R})\) have been proposed~\cite{26homotopy, 26homotopy2}. For completeness, here we show one of these methods~\cite{26homotopy} in detail. It suffices to prove the homotopy equivalence between \(\mathcal{A}(n)\) and Aff\(_{1}(n-1,\mathbb{R})\). Define a new space \(\mathcal{A}(n)\cap\mathrm{Aff}_{1}(n-1,\mathbb{R})\), which is exactly \(\mathcal{P}(n)\). Since we already proved \(\mathcal{A}(n)\simeq\mathcal{P}(n)\), it suffices to prove \(\mathrm{Aff}_{1}(n-1,\mathbb{R})\simeq\mathcal{P}(n)\). 

It is natural to define a map \(H_3\) on \(\mathrm{Aff}_{1}(n-1,\mathbb{R})\times [0,1]\) by
\begin{align}
H_3(G, t)\coloneqq \frac{1}{1-t\sum_ig_i}(G-t\ket{g}\bra{1}),
\end{align}
with \(g_i=\min(0,\min_{j}G_{ij})\leq0\). It is straightforward to verify \(\bra{1}H_3(G, t)=\frac{1}{1-t\sum_ig_i}(\bra{1}G-t\sum_ig_i\bra{1})=\frac{1}{1-t\sum_ig_i}(1-t\sum_ig_i)\bra{1}=\bra{1}\). Hence we have \(H_3(G, t)\in \mathrm{Aff}_{1}(n-1,\mathbb{R})\). 

It is straightforward to check \(H_3(G,0)=G\) and given \(G\in\mathcal{P}(n)\), we have \(H_3(G, t)=G\) due to \(g_i\equiv0\) for each \(i\). Moreover, we have \(\min(H_3(G,1))=\frac{\min_{i, j}(G_{ij}-g_i)}{1-\sum_ig_i}=\frac{\min_{i}\min_j(G_{ij}-g_i)}{1-\sum_ig_i}=\frac{\min_{i}(\min_jG_{ij}-g_i)}{1-\sum_ig_i}\geq0\) due to \(\min_jG_{ij}-g_i\geq0\) for each \(i\). Lastly, since the inequality \(1-t\sum_ig_i>0\) holds for \(t\in[0,1]\), we have 
\begin{align}
\begin{split}
&\det(H_3(G, t))\\
=&\frac{\det(G-t\ket{g}\bra{1}G)}{(1-t\sum_ig_i)^n}\\
=&\frac{\det(\mathbb{I}_n-t\ket{g}\bra{1})\det G}{(1-t\sum_ig_i)^n}\\
=&\frac{(1-t\braket{1|g})\det G}{(1-t\sum_ig_i)^n}\\
=&\frac{\det G}{(1-t\sum_ig_i)^{(n-1)}}\\
\neq&0. 
\end{split}
\end{align}
Here the third equality holds due to the Schur complement~\cite{12matrix}. Therefore, we have \(H_3(G,1)\in\mathcal{P}(n)\) and \(\mathcal{P}(n)\) is thus a strong deformation retract of Aff\(_{1}(n-1,\mathbb{R})\). 

In summary, \(\mathcal{A}(n)\) and \(\mathrm{GL}(n-1,\mathbb{R})\) have the same homotopy type. Since the topological classification is obtained in the limit \(n\to\infty\), this implies that the topological classification of stochastic systems with a non-zero point gap is exactly the same as that of quantum systems. 

\subsection{Hermitian case}
\label{sec:homotopyh}
\subsubsection{\(\mathcal{W}^{\mathrm{sym}}(n,-E_0)\simeq\mathcal{A}^{\mathrm{sym}}(n)\)}
Analogous to the non-Hermitian case in Sec.~\ref{sec:homotopynonh}, we have \(\mathcal{W}^{\mathrm{sym}}(n,-E_0)\simeq\mathcal{P}^{\mathrm{sym}}(n)\). However, in the Hermitian case, we cannot scale the column sum freely due to the additional constraint. Hence we have to use another method to prove further \(\mathcal{P}^{\mathrm{sym}}(n)\simeq\mathcal{A}^{\mathrm{sym}}(n)\). 

Intuitively, this homotopy means that a non-negative symmetric matrix can be continuously deformed to a doubly stochastic matrix~\cite{12matrix}, which inspires the idea of using Sinkhorn's theorem about positive symmetric matrices~\cite{66sinkhorn}. Hence, it is convenient to consider the interior of \(\mathcal{A}^{\mathrm{sym}}(n)\), i.e., \(\operatorname{int}{\mathcal{A}^{\mathrm{sym}}(n)}\coloneqq\{A\in\mathcal{A}^{\mathrm{sym}}(n)\mid A_{ij}>0,\forall i, j\}\). And we will prove that \(\operatorname{int}{\mathcal{A}^{\mathrm{sym}}(n)}\) and \(\mathcal{A}^{\mathrm{sym}}(n)\) have the same homotopy type. Given a matrix \(A\in\mathcal{A}^{\mathrm{sym}}(n)\), its singular values are continuous functions \(\sigma_1(A),\sigma_2(A),\dots,\sigma_n(A)\) in nonincreasing order with \(\sigma_i(A)>0\). Note that for any given \(n\times n\) matrices \(A_1\) and \(A_2\), we have~\cite{12matrix}
\begin{align}
|\sigma_n(A_1)-\sigma_n(A_2)|\leq\sigma_1(A_1-A_2).
\label{weyl}
\end{align}
Thus we can define a map \(\beta\) on \(\mathcal{A}^{\mathrm{sym}}(n)\) by \(\beta(A)\coloneqq A+A'\) with a positive symmetric matrix \(A'\coloneqq\frac{\sigma_n(A)}{2n}\ket{1}\bra{1}\). Since we have
\begin{align}
\begin{split}
|\sigma_n(\beta(A))-\sigma_n(A)|\leq&\sigma_1(A')\\
\sigma_n(\beta(A))\geq&\sigma_n(A)-\sigma_1(A')\\
\sigma_n(\beta(A))\geq&\sigma_n(A)-\sigma_n(A)\frac{\sigma_1(\ket{1}\bra{1})}{2n}\\
\sigma_n(\beta(A))\geq&\frac{\sigma_n(A)}{2}>0,
\end{split}
\label{weyl2}
\end{align}
\(\beta(A)\) is invertible and hence we have \(\beta(A)\in\operatorname{int}{\mathcal{A}^{\mathrm{sym}}(n)}\).

To prove \(\operatorname{int}{\mathcal{A}^{\mathrm{sym}}(n)}\simeq\mathcal{A}^{\mathrm{sym}}(n)\), we can construct a map \(H_4\) on \(\mathcal{A}^{\mathrm{sym}}(n)\times[0,1]\) by 
\begin{align}
H_4(A, t)\coloneqq A+tA'.
\end{align}
It is easy to check that \(H_4(A, t)\) satisfies the non-negativity constraint Eq.~\eqref{non-negativity} at each \(t\). Similarly, we have
\begin{align}
\begin{split}
|\sigma_n(H_4(A, t))-\sigma_n(A)|&\leq\sigma_1(A')\\
\sigma_n(H_4(A, t))&\geq (1-\frac{t}{2})\sigma_n(A)>0.
\end{split}
\end{align}
Therefore, \(H_4(A, t)\) is also invertible and thus we have \(H_4(A, t)\in\mathcal{A}^{\mathrm{sym}}(n)\). Furthermore, we have \(H_4(A,0)=A\) and \(H_4(A,1)=\beta(A)\in\operatorname{int}{\mathcal{A}^{\mathrm{sym}}(n)}\). Finally, for \(A\in\operatorname{int}{\mathcal{A}^{\mathrm{sym}}(n)}\), we have \(H_4(A, t)\in\operatorname{int}{\mathcal{A}^{\mathrm{sym}}(n)}\). Hence, \(\operatorname{int}{\mathcal{A}^{\mathrm{sym}}(n)}\) is a deformation retract of \(\mathcal{A}^{\mathrm{sym}}(n)\) in the weak sense~\cite{02algebraic}.

Similarly, we can show that \(\operatorname{int}{\mathcal{P}^{\mathrm{sym}}(n)}\) and \(\mathcal{P}^{\mathrm{sym}}(n)\) have the same homotopy type by constructing a map \(H_5\) on \(\mathcal{P}^{\mathrm{sym}}(n)\times[0,1]\) as 
\begin{align}
H_5(P, t)\coloneqq\frac{P+tP'}{1+\frac{t}{2}\sigma_n(P)}
\end{align} 
with \(P'\coloneqq\frac{\sigma_n(P)}{2n}\ket{1}\bra{1}\). Here the denominator is attached to satisfy probability conservation Eq.~\eqref{shiftedcon1} while other arguments remain the same. Therefore, we have \(\operatorname{int}{\mathcal{P}^{\mathrm{sym}}(n)}\simeq\mathcal{P}^{\mathrm{sym}}(n)\).

We start from a positive matrix \(A\in\operatorname{int}\mathcal{A}^{\mathrm{sym}}(n)\) and show that it can be continuously deformed to a doubly stochastic matrix in \(\operatorname{int}\mathcal{P}^{\mathrm{sym}}(n)\). Sinkhorn's theorem~\cite{66sinkhorn} states that given a positive symmetric matrix \(A\), there exists a unique positive diagonal matrix \(D_A\) such that \(D_AAD_A\) is doubly stochastic. Here \(D_A\) continuously depends on \(A\)~\cite{72sinkhorn}.

Then it is straightforward to prove \(\operatorname{int}\mathcal{P}^{\mathrm{sym}}(n)\simeq\operatorname{int}\mathcal{A}^{\mathrm{sym}}(n)\). More concretely, we define a map on \(\operatorname{int}\mathcal{A}^{\mathrm{sym}}(n)\times[0,1]\) by 
\begin{align}
H_6(A, t)\coloneqq D_A(t)AD_A(t)
\end{align}
with \(D_A(t)\coloneqq(1-t)\mathbb{I}_n+tD_A\). Since \(D_A\) has positive diagonal entries, so does \(D_A(t)\) and hence \(H_6(A, t)\in\operatorname{int}\mathcal{A}^{\mathrm{sym}}(n)\). Moreover, we can check \(H_6(A,0)=A\) and \(H_6(A,1)=D_AAD_A\in\operatorname{int}\mathcal{P}^{\mathrm{sym}}(n)\). For \(P\in\operatorname{int}\mathcal{P}^{\mathrm{sym}}(n)\), the unique \(D_A\) is the identity matrix \(\mathbb{I}_n\) and hence we have \(H_6(P, t)=P\). Therefore, \(\operatorname{int}\mathcal{P}^{\mathrm{sym}}(n)\) is a strong deformation retract of \(\operatorname{int}\mathcal{A}^{\mathrm{sym}}(n)\) and hence we have \(\mathcal{P}^{\mathrm{sym}}(n)\simeq\mathcal{A}^{\mathrm{sym}}(n)\).

\subsubsection{\(\mathcal{A}^{\mathrm{sym}}(n)\simeq\mathrm{GL}^{\mathrm{sym}}(n-1,\mathbb{R})\)}
It is argued that \(\mathcal{A}^{\mathrm{sym}}(n)\) and \(\mathrm{GL}^{\mathrm{sym}}(n-1,\mathbb{R})\) have the same homotopy type intuitively~\cite{26sym}. For completeness, here we show this method in detail.

We first show that \(\operatorname{int}\mathcal{A}^{\mathrm{sym}}(n)\) and \(\mathcal{C}^{\mathrm{sym}}(n-1)\coloneqq\{C\in\mathrm{GL}^{\mathrm{sym}}(n-1,\mathbb{R})\mid C_{ij}>-1\}\) have the same homotopy type. This means that the last column and the last row of \(A\) in \(\operatorname{int}\mathcal{A}^{\mathrm{sym}}(n)\) can be set to a fixed value. More concretely, we can construct a space \(\mathcal{A}^{\mathrm{sym}}_{n-1}(n)\coloneqq\{A\in\operatorname{int}\mathcal{A}^{\mathrm{sym}}(n)\mid A_{in}=A_{ni}=1,\forall i\}\) and define a map \(H_7\) on \(\operatorname{int}\mathcal{A}^{\mathrm{sym}}(n)\times [0,1]\) as 
\begin{align}
H_7(A, t)\coloneqq D_{\mathrm{last}}(t)AD_{\mathrm{last}}(t)
\label{Eq:sink}
\end{align}
where \(D_{\mathrm{last}}(t)\) is a family of diagonal matrices defined by \(D_{\mathrm{last}}(t)_i\coloneqq(\sqrt{A_{nn}}/A_{in})^t>0\) for \(i=1,\dots, n\) and is invertible. It is straightforward to verify \(H_7(A,0)=A\) and \((H_7(A,1))_{in}=(\sqrt{A_{nn}}/A_{in})A_{in}(\sqrt{A_{nn}}/A_{nn})=A_{in}/A_{in}=1\) for each \(i\). Moreover, for \(A\in\mathcal{A}^\mathrm{sym}_{n-1}(n)\), \(D_\mathrm{last}(t)=\mathbb{I}_n\). Therefore, \(\mathcal{A}^{\mathrm{sym}}_{n-1}(n)\) is a strong deformation retract of \(\operatorname{int}\mathcal{A}^{\mathrm{sym}}(n)\). Moreover, given \(A\in\mathcal{A}^{\mathrm{sym}}_{n-1}(n)\), it can be expressed as \(A=\begin{pmatrix}
A^{(n, n)}&\ket{1}_{(n-1)}\\
\bra{1}_{(n-1)}&1
\end{pmatrix}\) with an \((n-1)\times(n-1)\) positive symmetric matrix \(A^{(n, n)}\) and an all-ones \((n-1)\)-vector \(\ket{1}_{(n-1)}\). Then we have \(\det A =\det(A^{(n, n)}-\ket{1}_{(n-1)}\bra{1}_{(n-1)})\neq0\). If we define \(C\coloneqq A^{(n, n)}-\ket{1}_{(n-1)}\bra{1}_{(n-1)}\), the space \(\mathcal{C}^{\mathrm{sym}}(n-1)\) consisting of all possible \(C\)'s is homeomorphic to \(\mathcal{A}^{\mathrm{sym}}_{n-1}(n)\).

Then it suffices to prove that \(\mathcal{C}^{\mathrm{sym}}(n-1)\) is actually homotopy equivalent to \(\mathrm{GL}^{\mathrm{sym}}(n-1,\mathbb{R})\). For \(G\in\mathrm{GL}^{\mathrm{sym}}(n-1,\mathbb{R})\), we define a map \(\delta\) by \(\delta(G)\coloneqq \frac{G}{1+|\min G|}\) with \(\min G=\min_{ij} G_{ij}\). Since \(\det\delta(G)=\frac{1}{(1+|\min G|)^{n-1}}\det G\), \(\delta\) is a map to \(\mathcal{C}^{\mathrm{sym}}(n-1)\). If \(G\) is non-negative, we have \(\operatorname{min}\delta(G)=\frac{\operatorname{min}G}{1+\min G}\geq0\) and otherwise, we have
\begin{align}
\operatorname{min}\delta(G)&=-\frac{|\min G|}{1+|\min G|}>-1,
\end{align}
and thus \(\delta(G)\in\mathcal{C}^{\mathrm{sym}}(n-1)\).
Then we can define a map \(H_8\) on \(\mathrm{GL}^{\mathrm{sym}}(n-1,\mathbb{R})\times [0,1]\) by
\begin{align}
H_8(G, t)\coloneqq \frac{G}{1+t|\min G|}.
\end{align}
It is straightforward to verify \(H_8(G,0)=G\) and \(H_8(G,1)=\frac{G}{1+|\min G|}\equiv\delta(G)\in\mathcal{C}^\mathrm{sym}(n-1)\). Lastly, given \(C\in\mathcal{C}^{\mathrm{sym}}(n-1)\), we have \(\operatorname{det}H_8(C, t)=\frac{\operatorname{det}C}{(1+t|\min C|)^{(n-1)}}\neq0\) and \(\min H_8(C, t)=\frac{\min C}{1+t|\min C|}>-1\) similarly. Hence \(\mathcal{C}^{\mathrm{sym}}(n-1)\) is a deformation retract of \(\mathrm{GL}^{\mathrm{sym}}(n-1,\mathbb{R})\) in the weak sense.

Therefore, we prove that \(\mathcal{C}^{\mathrm{sym}}(n-1)\) and \(\mathrm{GL}^{\mathrm{sym}}(n-1,\mathbb{R})\) have the same homotopy type. This completes the proof of \(\mathcal{A}^{\mathrm{sym}}(n)\simeq\mathrm{GL}^{\mathrm{sym}}(n-1,\mathbb{R})\). 

\section{1D Topological classification in class AI}
\subsection{1D real-space invariant}
\label{sec:onedwind}
The corresponding real-space topological invariant in non-Hermitian 1D stochastic systems in class AI is defined as~\cite{24realspace}
\begin{align}
\begin{split}
	\nu_{\mathrm{real}}&\coloneqq \frac{1}{2}\operatorname{sig}((\widetilde{X}+i\widetilde{W^{\mathrm{sh}}})\widetilde{\Sigma})\\
&=\frac{1}{2}\operatorname{sig}\begin{pmatrix}
		X&-\mathrm{i}W^{\mathrm{sh}}\\\mathrm{i}(W^{\mathrm{sh}})^T&-X
	\end{pmatrix}.
\label{wind}
\end{split}
\end{align}
In translationally invariant systems in the thermodynamic limit \(n\to\infty\), this is equivalent to the \(\bm{k}\)-space winding number~\cite{26equivalence}
\begin{align}
\begin{split}
\nu&\coloneqq \oint\frac{dk}{2\pi\mathrm{i}}\partial_k\log\det(W(k)-E_0\mathbb{I}_n)\\
&=\oint\frac{dk}{2\pi\mathrm{i}}\partial_k\log\det A(k)\\
&=\frac{1}{2}\operatorname{sig}(M)
\end{split}
\end{align}
where \(A(k)\) obtained from \(W^\mathrm{sh}(k)\equiv W(k)-E_0\mathbb{I}_n\) homotopically is the Fourier form of a non-negative matrix as introduced in Sec.~\ref{sec:non-negative} and \(M\) is the mass term of \(A(k)\).

Analogous to the \(d\)-D topological classification of quantum systems~\cite{16classification}, the topological classification of \(d\)-D translationally invariant stochastic systems can be established by classifying \(\widetilde{M}\), obtained from the Dirac form \(\widetilde{A}(\bm{k})=\bm{k}\cdot\widetilde{\mathbf{\Gamma}}+\widetilde{M}\), with chiral symmetries \(\widetilde{\mathbf{\Gamma}}=(\widetilde{\Gamma}_1,\dots,\widetilde{\Gamma}_d)\) and momenta \(\bm{k}=(k_1,\dots, k_d)\). Here \(\widetilde{M}\) satisfies the same symmetries as \(\widetilde{A}\) besides additional chiral symmetries \(\widetilde{\Gamma}_i\). Thus we have to consider stochastic systems represented by a Hermitian mass term \(\widetilde{M}\in\mathcal{A}^{\mathrm{sym}}(2n)\) with chiral symmetries \(\widetilde{\Sigma}\) and \(\widetilde{\mathbf{\Gamma}}\).

Now we focus on the 1D case. Without loss of generality, we can choose a representation in which \(\widetilde{\Gamma}=\sigma_y\otimes \mathbb{I}_n\). Therefore, according to the anticommutation relations \(\{\widetilde{\Sigma},\widetilde{M}\}=0\) and \(\{\widetilde{\Gamma},\widetilde{M}\}=0\), we can determine the possible form of \(\widetilde{M}\) as
\begin{align}
\widetilde{M}=\sigma_x\otimes M.
\end{align}
Since \(\widetilde{M}\) is Hermitian, we have \(M=M^T\). Now the topology of \(\widetilde{M}\) depends on the topology of \(M\), which is actually a Hermitian non-negative matrix in class AI and 0D. Therefore, the topological classification of non-Hermitian 1D stochastic systems in class AI is \(\mathbb{Z}\) for transient states. 

\subsection{Why does a loop passing through zero with \(\nu=0\) vanish in the steady-state limit?}
\label{sec:nozeroloop}
Due to ergodicity, there is only one eigenvalue \(\lambda_0\) at 0. We can always find a unique loop \(\lambda_L(k)\) such that \(\lambda_L(0)=\lambda_0\). Note that since this loop can be composed of \(\lambda_i(k)\) with multiple possible \(i\)'s, we leave the index \(i\) unspecified. The fact that \(W\) is a real matrix ensures \begin{align}
W(k)=W^*(-k).
\label{Eq:real}
\end{align}
Hence given an eigenvalue \(\lambda_{\mathrm{L}}(k)\), there is another eigenvalue 
\begin{align}
\lambda_{\mathrm{L}}(2\pi-k)=\lambda_{\mathrm{L}}^*(k)
\label{Eq:conjugate}
\end{align}
in the spectrum. Since this loop corresponds to a gapped spectrum, in the vicinity of 0, there exists a momentum \(k_c<\pi\) such that \(\lambda_{\mathrm{L}}(k)\) possesses a non-zero imaginary part for \(k\in(0, k_c]\). According to Eq.~\eqref{Eq:conjugate}, \(\lambda_{\mathrm{L}}(k)\) is complex for \(k\in(0, k_c]\cup[2\pi-k_c,2\pi)\). Since \(\lambda_{\mathrm{L}}(k_c)\) and \(\lambda_{\mathrm{L}}(2\pi-k_c)\) have opposite imaginary parts, there must be at least one momentum \(k_r\in(k_c,2\pi-k_c)\) satisfying \(\lambda_{\mathrm{L}}(k_r)\in\mathbb{R}\) due to continuity. Without loss of generality, we can choose \(k_r\) to be the first such momentum as \(k\) increases from \(k_c\).

Next, we can divide the whole loop into two distinct loops on the intervals [\(0, k_r)\cup(2\pi-k_r,2\pi\)] and [\(k_r,2\pi-k_r\)] respectively. As \(E_0\to0^-\), the inequalities \(\lambda_{\mathrm{L}}(k_r)<E_0<0\) hold. Hence, the first loop possesses a non-zero winding number \(\nu=\pm1\). However, the second loop will not surround \(E_0\) in this limit and thus possesses a zero winding number. Consequently, the total winding number of the whole loop is always non-zero.

\subsection{The equivalence between two different \texorpdfstring{\(\bm{k}\)}{Lg}-space approaches in 1D}
\label{sec:tilt}
In this section, we will show that in 1D, our approach is equivalent to another tilting approach~\cite{17stochastic,24role}. For a 1D translationally invariant system, a tilted \(\bm{k}\)-space transition matrix \(W^{\lambda}(k)\) can be constructed from \(W^\lambda\). In particular, for a 1D stochastic system in the thermodynamic limit, the matrix entries of \(W^\lambda\) are defined as \(W^\lambda_{ab;\sigma\nu}\coloneqq W_{ab;\sigma\nu}e^{\lambda(a-b)}\) where \(a,b\) label unit cells and \(\sigma,\nu\) denote the internal degrees of freedom. Since \(W^\lambda\) no longer satisfies constraint~\eqref{Metzler}, the zero mode can be removed from the spectrum, allowing one to define a point gap away from zero. Accordingly, we have \(W^\lambda(k)=W(k-\mathrm{i}\lambda)\). The winding number of \(W^{\lambda}(k)\) is then calculated separately by taking two limits \(\lambda\to\pm0\) along positive and negative semi-axes respectively. The sum of these winding numbers is the topological invariant for the original transition matrix \(W(k)\). 

According to Eq.~\eqref{wind}, the singularity of the logarithm occurs at \(k_0\), the zero of the determinant \(\det(W(k)-E_0\mathbb{I}_n)\). When \(E_0\) approaches \(0^-\), we can expand the determinant as \(\det(W(k)-E_0\mathbb{I}_n)=\det{W(k)}-E_0\mathrm{tr}(\mathrm{adj} W(k))+\mathcal{O}(E_0^2)=k\partial_k\det W(0)-E_0\mathrm{tr}(\mathrm{adj} W(0))+\mathcal{O}(k^2, kE_0, E_0^2)\). Since \(\partial_k\det W(0)\neq0\) holds for gapped systems~\cite{24role}, up to leading order in \(k\) and \(E_0\), we have \(k_0=\frac{E_0(\mathrm{tr}\mathrm{adj} W(0))}{\partial_k\det W(0)}\). According to Eq.~\eqref{Eq:real}, we have \(\partial_k\det W(0)=-\partial_k\det W^*(0)=-(\partial_k\det W(0))^*\). Since \(W(0)\) is a real matrix, \(k_0\) is a purely imaginary number. Therefore, a shifting of a point gap \(E_0\) is equivalent to the shifting \(k\to k-\mathrm{i}\lambda\) for the winding number that is determined by the determinant of \(W(k)\).

More concretely, the winding number with respect to \(E_0\) is 
\begin{align}
\begin{split}
\nu=&\lim_{E_0\to0^-}\oint\frac{dk}{2\pi\mathrm{i}}\partial_k\log\det(W(k)-E_0\mathbb{I}_n)\\
=&PV\int^{\pi}_{-\pi}\frac{dk}{2\pi\mathrm{i}}\partial_k\log\det(W(k))\\
&+\lim_{E_0\to0^-}\int_{C(E_0)}\frac{dk}{2\pi\mathrm{i}}\partial_k\log\det(W(k)-E_0\mathbb{I}_n)\\
=&I_{PV}+\lim_{E_0\to0^-}\int_{C(E_0)}\frac{dk}{2\pi\mathrm{i}}\partial_k\log\det(W(k)-E_0\mathbb{I}_n)\\
=&I_{PV}+\lim_{E_0\to0^-}\int_{C(E_0)}\frac{dk}{2\pi\mathrm{i}}\frac{\partial_k\det{W(0)}}{k\partial_k\det{W(0)}-E_0(\mathrm{tr}\mathrm{adj} W(0))}\\
=&I_{PV}+\lim_{E_0\to0^-}\int_{C(E_0)}\frac{dk}{2\pi\mathrm{i}}\frac{1}{k-k_0}.
\label{minusE}
\end{split}
\end{align}
Here \(C(E_0)\) is a semicircle from \(k=r_c\) to \(k=-r_c\) centered at \(k=0\) with a small enough radius \(r_c\). As \(E_0\to0^-\), \(|k_0|\ll r_c\) holds. \(C(E_0)\) is located in the upper half-plane if \(\operatorname{sgn}(-\mathrm{i} k_0)>0\) and is located in the lower half-plane otherwise. 

Similarly, since the winding number is 0 if the point gap is set at \(-E_0\), we have
\begin{align}
\begin{split}
\nu_+=&\lim_{E_0\to0^-}\oint\frac{dk}{2\pi\mathrm{i}}\partial_k\log\det{(W(k)+E_0\mathbb{I}_n)}\\
=&I_{PV}+\lim_{E_0\to0^-}\int_{C(-E_0)}\frac{dk}{2\pi\mathrm{i}}\frac{1}{k+k_0}\\
=&I_{PV}-\lim_{E_0\to0^-}\int_{-C(E_0)}\frac{dk}{2\pi\mathrm{i}}\frac{1}{-k+k_0}\\
=&I_{PV}-\lim_{E_0\to0^-}\int_{C(E_0)}\frac{dk}{2\pi\mathrm{i}}\frac{1}{k-k_0}.
\label{plusE}
\end{split}
\end{align}
Here \(C(-E_0)\) is defined similarly for \(-E_0\), but is located in the opposite half-plane since the sign of the point gap is opposite. Note that since \(-E_0\) is not surrounded by the spectrum of \(W(k)\), we have \(\nu_+=0\). 

Adding Eqs.~\eqref{minusE} and \eqref{plusE} together, we have
\begin{align}
\begin{split}
\nu=2I_{PV}.
\end{split}
\label{Eq:shiftE0}
\end{align}

Such an argument can be applied to another approach. The winding number is given by \(w=w_++w_-\), where \(w_\pm=\lim_{\lambda\to0^\pm}\oint\frac{dk}{2\pi\mathrm{i}}\partial_k\log\det{(W(k-\mathrm{i}\lambda))}\). Similarly, we have
\begin{align}
\begin{split}
w&=2I_{PV}+\sum_{\epsilon=\pm1}\lim_{\lambda\to0^\epsilon}\int_{C(\lambda)}\frac{dk}{2\pi\mathrm{i}}\partial_k\log\det{(W(k-\mathrm{i}\lambda))}\\
&=2I_{PV},
\end{split}
\end{align}
which is exactly equal to Eq.~\eqref{Eq:shiftE0}.

This proves that the tilting approach is equivalent to our approach in 1D gapped cases. For gapless cases, the winding number in the tilting approach is always zero~\cite{24role}, rendering the phase trivial. In our approach, even though the winding number cannot be defined in this case, we still conclude that the phase with a gapless spectrum is trivial. In summary, we prove that these two approaches are generally equivalent in 1D.

Nevertheless, our approach offers greater computational efficiency since it requires only one limit rather than two limits in the tilted case where both \(\pm\lambda\) are supposed to be considered. Moreover, for higher \(d\), obtaining the topological invariant of \(W^\lambda(\bm{k})\) requires \(2^d\) limits. Therefore, defining the topological invariant directly from a non-zero point gap of \(W\) provides a significantly more efficient framework for the topological classification of stochastic systems, especially in higher dimensions.

Even though this method works well for 1D systems characterized by a single momentum \(k\), it becomes ambiguous when considering its higher-dimensional generalization with multiple momenta \(k_i\)'s. In particular, in one dimension, \(W^{\lambda}(k)\) can also be rewritten as \(W(k')\) with a shifted quasi-momentum \(k'=k-\mathrm{i}\lambda\). This naturally suggests that in higher dimensions, multiple tilting factors \(\lambda_i\) can be introduced to construct the shifted momenta \(k_i\)'s analogously. The shifted vector of momenta \(\bm{k}'\) is thus a multi-variable function of \(\{\lambda_i\}_{i=1,\dots, d}\). Consequently, there are infinitely many paths along which the possible limits can be taken. As a result, it remains unclear whether all these paths are equivalent and if not, which path is the correct one. More importantly, for 0D stochastic systems without the parameter \(\bm{k}\), it is impossible to define such a tilted transition matrix. Consequently, it is difficult to establish a comprehensive topological classification table in a unified framework considering only the tilted \(W^\lambda\).

\bibliographystyle{apsrev4-2}
\bibliography{ref}

\end{document}